\documentclass[twocolumn]{aastex61}
\usepackage{CJK}
\usepackage{amsmath}
\usepackage{verbatim}
\definecolor{blue}{RGB}{50, 80, 255}

\newcommand\aastex{AAS\TeX}

\shorttitle{\aastex\ Compositions of Planetary Debris around Dusty WDs}
\shortauthors{S. Xu et al.}

\begin{document}
\begin{CJK}{UTF8}{gbsn}

\title{Compositions of Planetary Debris around Dusty White Dwarfs}

\correspondingauthor{Siyi Xu}
\email{sxu@gemini.edu}
\author[0000-0002-8808-4282]{Siyi Xu (许\CJKfamily{bsmi}偲\CJKfamily{gbsn}艺)}
\affil{Gemini Observatory, 670 N. A'ohoku Place, Hilo, HI 96720}

\author{Patrick Dufour}
\affiliation{D$\acute{e}$partement de Physique, Universit$\acute{e}$ de Montr$\acute{e}$al, C.P. 6128, Succ. Centre-Ville, Montr$\acute{e}$al, Qu$\acute{e}$bec H3C 3J7, Canada}
\affiliation{Institut de Recherche sur les Exoplan$\grave{e}$tes (iREx), Universit$\acute{e}$ de Montr$\acute{e}$al, Montr$\acute{e}$al, QC H3C 3J7, Canada}

\author{Beth Klein} 
\affiliation{Department of Physics and Astronomy, University of California, Los Angeles, CA 90095-1562, USA}

\author{Carl Melis}
\affiliation{Center for Astrophysics and Space Sciences, UCSD, CA 92093-0424, USA}

\author{Nathaniel N. Monson}
\affiliation{Department of Earth, Planetary, and Space Sciences, University of California, Los Angeles, CA 90095, USA}

\author{B. Zuckerman}
\affiliation{Department of Physics and Astronomy, University of California, Los Angeles, CA 90095-1562, USA}

\author{Edward D. Young}
\affiliation{Department of Earth, Planetary, and Space Sciences, University of California, Los Angeles, CA 90095, USA}

\author{Michael A. Jura}
\altaffiliation{Deceased}
\affiliation{Department of Physics and Astronomy, University of California, Los Angeles, CA 90095-1562, USA}

\begin{abstract}
The photospheres of some white dwarfs are ``polluted" by accretion of material from their surrounding planetary debris. White dwarfs with dust disks are often heavily polluted and high-resolution spectroscopic observations of these systems can be used to infer the chemical compositions of extrasolar planetary material. Here, we report spectroscopic observation and analysis of 19 white dwarfs with dust disks or candidate disks. The overall abundance pattern very much resembles that of bulk Earth and we are starting to build a large enough sample to probe a wide range of planetary compositions. We found evidence for accretion of Fe-rich material onto two white dwarfs as well as O-rich but H-poor planetary debris onto one white dwarf. In addition, there is a spread in Mg/Ca and Si/Ca ratios and it cannot be explained by differential settling or igneous differentiation. The ratios appear to follow an evaporation sequence. In this scenario, we can constrain the mass and number of evaporating bodies surrounding polluted white dwarfs.
\end{abstract}


\section{Introduction}

Thanks to radial velocity, transit, and microlensing surveys, we now know that planetary systems are prevalent around main-sequence stars \citep[e.g.][]{Petigura2013}. Naturally, this has lead to great interest in performing detailed characterization of extrasolar planetary systems. Using mass-radius relationships, we can differentiate between rocky planets, ocean planets, and planets with an extended atmosphere \citep{Zeng2016}. However, planet models become increasingly degenerate with little difference in planet mass-radius relationships when including minor and trace elements \citep[e.g.][]{Dorn2015}. Luckily, we can get around this limitation by observing ``polluted" white dwarfs (WDs), which are accreting from extrasolar planetary debris. 

Recent studies show that planetary systems can be present around WDs \citep{MustillVillaver2012,NordhausSpiegel2013}. From different dynamical interactions, i.e. planet-planet scattering, mean motion resonance, and Kozai-Lidov effect, minor planets can be perturbed to enter into the tidal radius of the WD and subsequently be disrupted \citep[e.g.][]{DebesSigurdsson2002,FrewenHansen2014,Stephan2017}. Eventually, all these planetary debris will settle into a disk of dust and gas. Dust disks around WDs can be detected via an infrared excess because most of the WD light is in the UV/optical \citep[e.g.][]{Jura2007b}. Circumstellar gas material can manifest itself as either emission or absorption, depending on the inclination of the disk \citep[e.g.][]{Gaensicke2006, Gaensicke2012, Debes2012b}.

Thereafter, the circumstellar dust and gas will be accreted onto the WD, polluting its atmosphere. Because the outermost layer of WDs typically consist of pure hydrogen (DA) or helium (DB), even a small amount of heavy elements can create a detectable spectroscopic signature and the constituent elements can be measured individually to a high precision \citep[e.g.][]{Zuckerman2007}. Spectroscopic studies of these polluted WDs can be used to infer the bulk compositions of extrasolar planetary material, which is not possible with any other technique \citep{JuraYoung2014}.

A direct support for this asteroid disruption model is the discovery of an actively disintegrating asteroid transiting WD~1145+017 near the tidal radius \citep{Vanderburg2015}. Recently, it has been suggested that there is a planetesimal in a 2-hr orbit around WD~1226+110 \citep{Manser2019}. Not surprisingly, both WD~1145+017 and WD~1226+110 have a circumstellar dust and gas disk and they are heavily polluted \citep{Gaensicke2012,Brinkworth2012,Xu2016}.

To zeroth order, the compositions of extrasolar planets accreted onto WDs resemble that of bulk Earth -- O, Fe, Si, and Mg are often the dominant elements \citep{Xu2014}. Planets with exotic compositions that have no solar system analogs, such as refractory-rich planets or carbon-rich planets, have yet to be found \citep{JuraXu2013, Wilson2016}. In a volume-limited sample of DBs, the fraction of water is less than 1\% of the total accreted mass \citep{JuraXu2012} even though there are some exceptions where the WDs have accreted a significant amount of water \citep[e.g.][]{Farihi2013, Raddi2015, Gentile-Fusillo2017}. At least one WD has accreted a Kuiper-Belt-Object analog with 30\% water and 10\% carbon by mass \citep{Xu2017}. Differentiation and collisions also appear to be common in extrasolar planetary systems \citep{Jura2013b, Harrison2018}.

Identifying and analyzing heavily polluted WDs \textcolor{blue}{is} crucial to our understanding of the chemical compositions of extrasolar planets. Even though hundreds of polluted WDs have been identified, most of them only show calcium absorption in the optical \citep{Zuckerman2003,Koester2005,Dufour2007,Hollands2018}. The most heavily polluted WDs often display excess infrared radiation from an orbiting dust disk \citep{vonHippel2007}. Dusty WDs are promising targets to detect multiple elements -- the focus of this study. 

In this paper, we report results from high resolution spectroscopic observations of 13 WDs with circumstellar dust disks and 6 disk candidates. Due to different false positives, WDs with infrared excesses are considered as disk candidates in this work unless a {\it Spitzer} observation exists to confirm the disk nature of the infrared excess. In this sample, 16 WDs are DAs, 2 are DBs, and 1 is a DZ\footnote{Based on the spectroscopic classification, WD~1232+563 is a DZBA because the strongest optical absorption feature is Ca II-K. }. Most of the observations were performed at the {\it Keck Telescope}; one southern object was observed with the {\it Very Large Telescope} ({\it VLT}) and another system with the {\it Hubble Space Telescope} ({\it HST}). The paper is organized as follows. Observations and data reduction are described in Section~\ref{sec:obs}. Abundance analysis is presented in Section~\ref{sec:abund} and the results are discussed in Section~\ref{sec:discussion}. Conclusions are presented in Section~\ref{sec:conclusion}.

\section{Observations \label{sec:obs}}

\subsection{Keck/HIRES}

The majority of WDs in our sample were observed with the High Resolution Echelle Spectrometer (HIRES) on the {\it Keck I Telescope} \citep{Vogt1994}, which has a blue collimator (HIRESb) and a red collimator (HIRESr). For both collimators, the C5 decker was used, which gives a spectral resolution of $\sim$ 40,000. The typical wavelength coverage is 3200--5750~{\AA} and 4700--9000~{\AA} for HIRESb and HIRESr, respectively. Following our previous HIRES observations, data reduction was performed by using the MAKEE package and then continuum normalization with IRAF \citep{Klein2010,Xu2016}. The observing log and some representative spectra are shown in the Appendix.

\subsection{Keck/ESI}

Some of the WDs were observed with the Echellette Spectrograph and Imager (ESI) on the {\it Keck II Telescope} \citep{Sheinis2002}. A slit width of 0{\farcs}3 was used with a spectral resolution of 14,000. The wavelength coverage is 3900--10,000~{\AA}. Data reduction was performed with MAKEE and IRAF, similar to the HIRES reduction process.

Compared to HIRES, ESI has a lower spectral resolution but a wider wavelength coverage. ESI is more efficient at studying faint heavily polluted WDs, e.g. WD~1232+563. However, ESI has no transmission shortward of 3900~{\AA} and the throughput is very poor around Ca II K 3933 -- the most important absorption line in the optical. HIRESb is crucial for detecting lines between 3200 and 4000~{\AA}.

\subsection{VLT/UVES}

WD 0107-192 was observed with the Ultraviolet and Visual Echelle Spectrograph (UVES) on the {\it VLT} under program 096.C-0132 \citep{Dekker2000}. The dichroic beam splitter was used, which allows for simultaneous observations with both the blue and the red arms. For the blue arm, the CD\#2 grating was used with a wavelength coverage of  3300--4500~{\AA} while for the red arm, the CD\#4 grating was used with a wavelength coverage of 5700--9400~{\AA}. A slit width of 1{\farcs}0 was chosen, which gives a spectral resolution of 22,000. Data reduction was performed using the UVES pipeline.

\subsection{HST/COS}

PG~0010+280 is the hottest dusty WDs known and optical observations with HIRES only returned upper limits \citep{Xu2015}. Here, we report UV spectroscopic observations with the Comic Origins Spectrograph (COS) onboard the {\it HST} (Program ID \#14117). The G130M grating was used with a central wavelength of 1291~{\AA}, which gives a wavelength coverage of 1150--1430~{\AA} with a 20~{\AA} gap in the middle. To minimize the fixed-pattern noise, all four FP-POS steps are used (COS Instrument Handbook). Data reduction was performed with the CALCOS pipeline.

\section{Data Analysis  \label{sec:abund} }

\subsection{Stellar Parameters \label{sec:WDpar}}

\begin{deluxetable}{llllllllll}
\tablecaption{WD Parameters  \label{tab:par}}
\tablewidth{0pt}
\tablehead{
 \colhead{Name} & \colhead{Atm.} & \colhead{T (K) } & \colhead{Log g} & Ref \tablenotemark{a}
}
\startdata
G 166-58 & H	&7390 $\pm$ 200 & 7.99 $\pm$ 0.10 & (1,2)	\\
WD 2221-165 & H	 & 10130 $\pm$ 200 & 8.15 $\pm$ 0.10 &  (3,4)	\\
WD 0307+077 & H	 & 10230 $\pm$ 200& 7.96 $\pm$ 0.10	&  (3,4)  \\
PG 1541+651 & H	 &11880 $\pm$ 200 & 8.20 $\pm$ 0.10 &  (1,5)	\\
WD 1145+288\tablenotemark{b}	& H	&12140 $\pm$ 210 & 8.14 $\pm$ 0.10&  {(3,6)} \\
WD 1150-153 & H &12640 $\pm$ 200	& 8.22 $\pm$ 0.10 & (1,7)\\
GD 56	& H		 &15270 $\pm$ 300	& 8.09 $\pm$ 0.10	& (1,8)	\\
WD 0107-192\tablenotemark{b}	& H& 15440 $\pm$ 300	& 7.95 $\pm$ 0.10 &(1,9) \\
HE 0106-3253 & H	& 17350 $\pm$ 200& 8.12 $\pm$ 0.10	& (3,4)	\\
PG 1015+161 & H	&20420	$\pm$ 350 & 8.11 $\pm$ 0.10 &(1,8)	\\
PG 1457-086\tablenotemark{b,c} & H	&22240 $\pm$ 400  & 7.99 $\pm$ 0.10 &   (1,9,10)	\\
WD 1226+110 & H	 &23500 $\pm$ 200	& 8.16 $\pm$ 0.10 & (11,12) 	\\
PG 1018+411 & H	&24440 $\pm$ 400 & 8.11 $\pm$ 0.10 & (1,13)	\\
PG 0843+517 & H	 & 24670 $\pm$ 400	& 7.93 $\pm$ 0.10 & (1,14)	\\
WD 1341+036\tablenotemark{b} &H	& 26420 $\pm$ 200 & 7.85 $\pm$ 0.10 &(11,15)	\\
PG 0010+280\tablenotemark{b}	& H	& 27220 $\pm$ 400	& 7.87 $\pm$ 0.10 & (1,16)	\\
\\
WD 1232+563\tablenotemark{b} &He	&11787 $\pm$ 423\tablenotemark{b} & 8.30 $\pm$ 0.06 &  (17) \\
WD 1551+175 & He	& 14756 $\pm$ 1286 & 8.02 $\pm$ 0.12\tablenotemark{b} &  (3,18)\\
WD 2207+121 & He	& 14752 $\pm$ 1192 &  7.97 $\pm$ 0.12\tablenotemark{b} &(17,14)\\
\enddata
\tablecomments{(1) \citet{Gianninas2011}; (2) \citet{Farihi2008b}; (3) {\it This paper}; (4) \citet{Farihi2010b}; (5) \citet{Kilic2012}; (6) \citet{Barber2014}; (7) \citet{KilicRedfield2007}; (8) \citet{Jura2007b}; (9) \citet{Dennihy2017}; (10) \citet{Farihi2009}; (11) \citet{Tremblay2011}; (12) \citet{Brinkworth2009}; (13) \citet{Barber2016}; (14) \citet{XuJura2012}; (15) \citet{Li2017}; (16) \citet{Xu2015}; (17) \citet{Coutu2019}; (18) \citet{Bergfors2014}}
\tablenotetext{a}{The first reference is for the WD parameters and the second is for the infrared excess.}
\tablenotetext{b}{Disk candidates.}
\tablenotetext{c}{This object has {\it Spitzer} photometry. However, there is a red background object within 1 arcsec and the photometry is likely to be contaminated \citep{Dennihy2017}.}
\end{deluxetable}

The stellar parameters for most DAs are taken from previous studies using a spectroscopic method \citep{Gianninas2011, Tremblay2011} -- it compares the profiles of normalized Balmer lines with WD models until a good match is found. The uncertainties quoted in those works are only statistical uncertainties from the fit, which could underestimate the true uncertainties. Here, we adopt a more conservative uncertainty of at least 200~K in temperature and 0.1 dex in log~g. For WDs not in these two studies, we performed our own spectroscopic fit with the {\it SDSS} spectra or other archival spectra using models from the Montreal white dwarf group following the same procedures outlined in these two papers. Our adopted parameters, after applying 3D corrections \citep{Tremblay2011}, are listed in Table~\ref{tab:par}.

The analysis of DBs and DZs are a more interactive process. Following the methods developed in \citet{Dufour2007} and \citet{Coutu2019}, we started with finding the best WD parameters that would fit the {\it SDSS} photometry and parallax from {\it Gaia} while also including hydrogen and heavy elements. Then we used this set of parameters as the starting point to do an interactive fitting of the {\it SDSS} spectrum by varying temperature, log g, H and Ca abundances\footnote{The other heavy elements are included assuming the composition of CI chondrites.}. This whole process is repeated several times until the parameters have converged to a stable solution (see \citealt{Coutu2019} for details). The derived WD parameters are listed in Table~\ref{tab:par}.

\subsection{Abundance Determination \label{sec:AbundanceDetermination}}

\begin{deluxetable*}{lllllllllll}
\tablecaption{Abundances of DAZs \label{tab:DAZ}}
\tablewidth{0pt}
\tablehead{
\colhead{Name} & \colhead{Velocity}  & \colhead{log n(Mg)/n(H)} & \colhead{log n(Si)/n(H)} & \colhead{log n(Ca)/n(H)} & \colhead{log n(Fe)/n(H)} & \colhead{log n(Z)/n(H)} \\
& (km s$^{-1}$) & 
}
\startdata
G 166-58 & 29.1 $\pm$ 1.0	& -8.06 $\pm$ 0.05 &  $<$ -8.20	& -9.33 $\pm$ 0.08 & -8.22 $\pm$ 0.13 &	Ni -9.50 $\pm$ 0.20\\
WD 2221-165 & 45.9 $\pm$ 1.0& -7.00 $\pm$ 0.20	&  $<$ -5.00 &  -7.52 $\pm$ 0.15 &  $<$ -5.90 & ...\\
WD 0307+077 & 98.4 $\pm$ 1.1 & -6.58 $\pm$ 0.07	&  $<$ -5.50	& -7.10 $\pm$ 0.14	&  $<$ -6.30 & ...\\
PG 1541+651 & 102 $\pm$ 1.0	&  $<$ -6.50	&  $<$ -5.70	& -7.36 $\pm$ 0.08	&  $<$ -6.00 & ...\\
WD 1145+288	& 41.7 $\pm$ 1.5	& -6.00 $\pm$ 0.20	&  $<$ -4.70	& -6.88 $\pm$ 0.08	& -5.43 $\pm$ 0.20 & ...\\
WD 1150-153 & 21.8 $\pm$ 1.8	& -6.14 $\pm$ 0.20	&  $<$ -5.50	& -7.03 $\pm$ 0.20	&  $<$ -5.70	& ... \\
GD 56		& 19.5 $\pm$ 2.1	& -5.55 $\pm$ 0.20 & -5.69 $\pm$ 0.20 & -6.86 $\pm$ 0.20 &  -5.44 $\pm$ 0.20  & ... \\
WD 0107-192	&33.6 $\pm$ 2.0 &	 $<$ -6.40 &  $<$ -5.20 & -7.77 $\pm$ 0.20&  $<$ -4.70 & ...\\
HE 0106-3253 	& 55.9 $\pm$ 1.0 & -5.57 $\pm$ 0.20 & -5.48 $\pm$ 0.05 &-5.93 $\pm$ 0.11 & -4.70 $\pm$ 0.06 &  ...\\
PG 1015+161 & 68.2 $\pm$ 1.0 &	-5.60	 $\pm$ 0.20	&-5.42 $\pm$ 0.21	& -6.40 $\pm$ 0.20 & -4.92 $\pm$ 0.20 & ... \\
PG 1457-086 & 22.6	$\pm$ 1.0 & -5.47 $\pm$ 0.20	& -5.85 $\pm$ 0.20 & -6.23 $\pm$ 0.20 & $<$ -5.0 & ...\\
WD 1226+110 & 38.5 $\pm$ 2.5 & -4.52 $\pm$ 0.20 & -4.64 $\pm$ 0.15 & -5.26 $\pm$ 0.16 &  $<$ -4.50	& ...\\
PG 1018+411 & 30.5 $\pm$ 2.0 & -4.86 $\pm$ 0.20 & -5.36 $\pm$ 0.20 &  $<$ -5.00  &  $<$ -4.30 & ... \\
PG 0843+517 & 81.1 $\pm$ 3.1	&-4.82 $\pm$ 0.20 & -4.59 $\pm$ 0.12 & -6.26 $\pm$ 0.20 & -3.84 $\pm$ 0.18	& ...\\
WD 1341+036 & ...	&  $<$ -6.00	&  $<$ -5.50	&  $<$ -4.50 &  $<$ -4.50 & ...\\
PG 0010+280		& 40.9 $\pm$ 5.8	& ...	& -7.51 $\pm$ 0.15 & ...& ...& C -6.75/-8.01\tablenotemark{a} $\pm$ 0.15 \\
\enddata
\tablenotetext{a}{The carbon abundance of -6.75 is derived from using C II lines while -8.01 is from C III lines.}
\tablecomments{Spectral lines used for abundance analysis: Mg I 3832.3~{\AA}, 3838.3~{\AA}, 5172.7~{\AA}, 5183.6~{\AA}, Mg II 4481 doublet, Si II 3856.0~{\AA}, 3862.6~{\AA}, 4128.1~{\AA}, 4130.9~{\AA}, 5041.0~{\AA}, 5056.0~{\AA}, 6347.1~{\AA}, 6371.4~{\AA}, Ca I 4226.7~{\AA}, Ca II 3158.9~{\AA}, 3179.3~{\AA}, 3706.0~{\AA}, 3736.9~{\AA}, 3933.7~{\AA}, 3968.5~{\AA}, 8498.0~{\AA}, 8542.1~{\AA}, 8662.1~{\AA}, Fe I 3734.9~{\AA}, 3859.9~{\AA}, Fe II 3213.3~{\AA}, 3227.7~{\AA}, 4233.2~{\AA}, 4923.9~{\AA}, 5018.4~{\AA}, 5169.0~{\AA}. For G166-58, we also used many Fe I lines between 3400--3900~{\AA}. For PG 0010+280, we used C II 1334.5~{\AA}, 1335.7~{\AA}, many C III lines around 1175~{\AA}, Si II 1260.4~{\AA}, 1264.7~{\AA}, Si III 1294.5~{\AA}, 1296.7~{\AA}, 1298.9~{\AA}, Si IV 1393.8~{\AA}, 1402.8~{\AA}.
}
\end{deluxetable*}

\begin{deluxetable*}{lccccccccc}
\tablecaption{Accretion Rates of DAZs \label{tab:DAZacc}}
\tablewidth{0pt}
\tablehead{
\colhead{Name} & \colhead{Mg (g s$^{-1}$)} & \colhead{Si (g s$^{-1}$)} & \colhead{Ca (g s$^{-1}$)} & \colhead{Fe (g s$^{-1}$)} 
}
\startdata
G 166-58  & 5.6$\times$ 10$^6$ & $<$4.6$\times$10$^6$ & 5.2$\times$10$^5$ & 1.3$\times$10$^7$\\
WD 2221-165 & 2.7$\times$ 10$^7$ & $<$3.0$\times$10$^9$ & 1.3$\times$10$^7$ & $<$1.1$\times$10$^9$ \\
WD 0307+077 & 4.5$\times$ 10$^7$ & $<$6.1$\times$10$^8$ & 2.2$\times$10$^7$ &  $<$ 2.7$\times$10$^8$  \\
PG 1541+651 &   $<$1.9$\times$10$^7$ & $<$8.1$\times$10$^7$ & 4.1$\times$10$^6$ & $<$ 1.4$\times$10$^8$  \\
WD 1145+288	& 3.2$\times$10$^7$ & $<$4.0 $\times$10$^8$ & 7.1$\times$10$^6$ & 2.8$\times$10$^8$  \\
WD 1150-153  & 1.0$\times$10$^7$ & $<$4.3 $\times$10$^7$  & 2.7$\times$10$^6$ & $<$1.0$\times$10$^8$ \\
GD 56		& 1.0$\times$10$^7$ &  1.3$\times$10$^7$ & 1.3$\times$10$^6$ & 1.1$\times$10$^9$ \\
WD 0107-192	& $<$1.2$\times$10$^6$ & $<$3.4$\times$10$^7$ & 1.3$\times$10$^5$  & $<$3.9$\times$10$^8$ \\
HE 0106-3253 	 & 8.9$\times$ 10$^6$ & 2.0$\times$10$^7$  & 9.9$\times$10$^6$ & 4.4$\times$10$^8$ \\
PG 1015+161 & 9.3$\times$ 10$^6$ & 2.1$\times$10$^7$  & 4.0$\times$10$^6$ & 2.5$\times$10$^8$ \\
PG 1457-086 & 1.2$\times$ 10$^7$ & 6.8$\times$10$^6$  & 5.9$\times$10$^6$ &$<$9.2$\times$10$^7$  \\
WD 1226+110 & 1.4$\times$ 10$^7$ & 1.4$\times$10$^8$ & 6.9$\times$10$^7$& $<$7.2$\times$10$^8$  \\
PG 1018+411  & 6.3$\times$ 10$^7$ & 2.4$\times$10$^7$ & $<$1.2$\times$10$^8$  & $<$1.1$\times$10$^9$ \\
PG 0843+517 & 5.7$\times$ 10$^7$ & 1.1$\times$10$^8$ & 5.7$\times$10$^6$ & 2.4$\times$10$^{9}$\\
WD 1341+036 & $<$ 3.7$\times$ 10$^6$ & $<$1.2$\times$10$^7$ & $<$3.3$\times$10$^8$  & $<$4.4$\times$10$^8$ \\
PG 0010+280  & ... & 1.1$\times$ 10$^5$ & ..& ....\\
\enddata
\tablecomments{The accretion rates are calculated based on abundances reported in Table~\ref{tab:DAZ} and diffusion timescales from the Montreal White Dwarf Database \citep{Dufour2017}.
}
\end{deluxetable*}

\begin{deluxetable*}{lccccccccccc}
\tablecaption{Abundances and Accretion Rates of DBZs and DZs \label{tab:DBZ}}
\tablewidth{0pt}
\tablehead{
& \multicolumn{3}{c}{WD 1232+563} &   \multicolumn{3}{c}{WD 1551+175}  &  \multicolumn{3}{c}{WD 2207+121}\\
& log n(Z)/n(He) & M(Z)& $\dot{M}(Z)$ &  log n(Z)/n(He)  & M(Z) (g) & $\dot{M}(Z)$ & log n(Z)/n(He)  & M(Z) (g) & $\dot{M}(Z)$  \\
& &  (10$^{20}$g) &  (10$^8$ g s$^{-1}$)  & & (10$^{20}$g) &  (10$^8$ g s$^{-1}$)  & & (10$^{20}$g) &  (10$^8$ g s$^{-1}$)  
}
\startdata
H	& -5.90 $\pm$ 0.15 & 5.80 &...	& -4.45 $\pm$ 0.08 & 219.47 &... & -6.32 $\pm$ 0.15 & 3.52 & ...\\
O & -5.14 $\pm$ 0.15 & 528.29 & 33.57 & -5.48 $\pm$ 0.15 & 323.60 & 11.89 & -5.32 $\pm$ 0.15 & 557.96 & 16.43\\
Mg & -6.09 $\pm$ 0.05 & 89.63 &5.47 & -6.29 $\pm$ 0.05 & 75.79 & 2.77 & -6.15 $\pm$ 0.10 & 126.24 & 3.71\\
Al &  $<$ -7.50	& $<$ 3.91 & $<$ 0.24 & -6.99 $\pm$ 0.15 & 16.90 	&0.64 & -7.08 $\pm$ 0.15 & 16.58 & 0.50\\
Si & -6.36 $\pm$ 0.13 & 56.53 & 3.40 & -6.33 $\pm$ 0.10	& 79.87 & 3.00 & -6.17 $\pm$ 0.11& 139.95 & 4.23\\
Ca & -7.69 $\pm$ 0.05 & 3.77 & 0.38 & -6.93 $\pm$ 0.07	& 28.50 & 1.59 & -7.40 $\pm$ 0.08 & 11.65 & 0.52\\
Ti & -8.96 $\pm$ 0.11 & 0.24 & 0.03 & -8.68 $\pm$ 0.11	& 0.62 & 0.04 &  -8.84 $\pm$ 0.14 & 0.51 & 0.02\\
Cr & -8.16 $\pm$ 0.07 & 1.64 & 0.17 & -8.25 $\pm$ 0.07	& 1.80 & 0.11 & -8.16 $\pm$ 0.19 & 2.62 & 0.12\\
Mn & -8.54 $\pm$ 0.05 & 0.73 & 0.08 & -8.74 $\pm$ 0.05 	& 0.61 & 0.04 & -8.50 $\pm$ 0.08 & 1.26 & 0.06\\
Fe & -6.45 $\pm$ 0.11 & 90.73 & 9.18 & -6.60 $\pm$ 0.10	& 85.68 & 4.85 & -6.46 $\pm$ 0.13 & 142.72 & 6.48\\
Ni & $<$ -7.30 & $<$ 13.47 & $<$ 1.28 & $<$ -7.50 &  $<$ 11.34 & $<$ 0.62 &-7.55 $\pm$ 0.20 & 12.11 & 0.53\\
$\sum$\tablenotemark{a} &... & 771.55 & 52.50 & ... & 613.38 & 24.92 & ... & 1011.61 & 32.61 \\
\enddata
\tablenotetext{a}{The total mass and accretion rate exclude H.}
\tablecomments{The average radial velocity is 19.0 $\pm$ 2.0 km s$^{-1}$, 22.9 $\pm$ 1.5 km s$^{-1}$, and 34.5 $\pm$ 2.0 km s$^{-1}$ for WD~1232+563, WD~1551+175 and WD~2207+121, respectively.
}
\end{deluxetable*}

To determine the abundances of heavy elements, we calculated synthetic spectra using the WD atmospheric structure computed in section~\ref{sec:WDpar}. In each iteration, a grid of spectra with different abundances for the element of interest were calculated while keeping the abundances of all the other elements the same. Each model spectrum was compared with the data and the final abundance is adopted from the model returning the minimum $\chi^2$. This process is repeated for each heavy element for all the DAZs. Some representative fits are shown in the Appendix.

If only one absorption line is detected for a heavy element, the abundance uncertainty mostly comes from the quality of the spectrum. Most elements have more than one detected line and the uncertainty is dominated by the different abundances derived when using different spectral lines. In addition, to estimate the impact of stellar parameters, we computed abundances of heavy elements with a five-grid WD model with parameters of (T, log g), (T+$\Delta$T, log g),  (T-$\Delta$T, log g), (T, log g + $\Delta$ log g), and (T, log g - $\Delta$ log g). The final abundances and uncertainties listed in Table~\ref{tab:DAZ} are the average values from these five models. The accretion rates are listed in Table~\ref{tab:DAZacc}. We consider the abundance uncertainties to be fairly conservative, as they include both the measurement error and systematic error from WD parameters.

For DBZs and DZs, the presence of heavy elements could also have an impact on the thermal structure of a WD model. Sometimes, it is necessary to repeat the processes outlined in section~\ref{sec:WDpar} to obtain a self-consistent solution. It has been demonstrated that this method works well even for heavily blended regions, e.g. WD~J0738+1835 \citep{Dufour2012}. Some representative fits are shown in the Appendix and the abundances and accretion rates are reported in Table~\ref{tab:DBZ}.

{\it PG 0010+280} is the only WD with UV spectra in this sample. Our model can reasonably reproduce the strength of Si~III and Si~IV lines, as shown in the Appendix. However, we found a big discrepancy of carbon abundances derived from C~II and C~III, likely because the atmosphere is stratified due to gravitational settling and radiative levitation \citep{Koester2014a}. The equivalent width of Si II 1265~{\AA} and C II 1335~{\AA} lines are 34~m{\AA} and 81~m{\AA}, respectively; this is within the range that can be supported by radiative levitation and no ongoing external accretion is required to explain the heavy elements in the atmosphere \citep{Koester2014a}.

\subsection{Optical and UV Discrepancy \label{sec:discrep}}

The abundances of three DAZs (PG~1015+161, WD~1226+110, and PG~0843+517) analyzed here have been reported in a previous study with UV spectroscopy \citep{Gaensicke2012}. We find some discrepancies between optical and UV abundances, as listed in Table~\ref{tab:optUV}. In these three DAZs, the optical Si and Fe abundances are always higher than those derived from the UV data. The most extreme case is Si in PG~1015+161, where the optical abundance is a factor of 10 higher than that in the UV. We have identified five other systems from the literature where optical and UV determinations of the same element exist. For two of the DBZs (GD~40 and G~241-6), their optical Si abundances are lower than the UV abundances -- opposite of what we found here for the DAZs. The Si abundances in the other two DBZs (GD~61 and Ton~345) agree within the uncertainties between the optical and UV.

\begin{deluxetable*}{lccccccccc}
\tablecaption{Optical and UV Abundance Comparison \label{tab:optUV}}
\tablewidth{0pt}
\tablehead{
\colhead{Name} &\colhead{Atm} & \colhead{T (K)} & \colhead{log g} & \colhead{log n(C)/n(H(e))} & \colhead{log n(O)/n(H(e))} & \colhead{log n(Si)/n(H(e))} & \colhead{log n(Fe)/n(H(e))}  & \colhead{Ref}
}
\startdata
PG 1015+161\tablenotemark{o} & H & 20420 & 8.11 & ... & ... & -5.42 $\pm$ 0.21 & -4.92 $\pm$ 0.20 & {\it this paper}\\
PG 1015+161\tablenotemark{uv} & H & 19200 & 8.22 &$<$ -8.00 &  -5.50 $\pm$ 0.20 &  -6.40 $\pm$ 0.20 & -5.50 $\pm$ 0.30 &  \citet{Gaensicke2012} \\
WD 1226+110\tablenotemark{o} & H & 23500 & 8.16 & ... & ... & -4.64 $\pm$ 0.15 & $<$ -4.50 & {\it this paper} \\
WD 1226+110\tablenotemark{uv} & H & 20900 & 8.15 & - 7.50 $\pm$ 0.20 & -4.55 $\pm$ 0.20 & -5.20 $\pm$ 0.20 & -5.20 $\pm$ 0.30 & \citet{Gaensicke2012} \\
WD 1929+012\tablenotemark{o} & H & 20890 & 7.90 & $<$ -4.15 & -3.62 $\pm$ 0.05  & -4.24 $\pm$ 0.07 & -4.43 $\pm$ 0.09& \citet{Vennes2010} \\
WD 1929+012\tablenotemark{o} & H & 23470 & 7.99 &$<$ -4.85 & -3.68 $\pm$ 0.10 & -4.35 $\pm$ 0.11& -4.10 $\pm$ 0.10& \citet{Melis2011} \\
WD 1929+012\tablenotemark{uv} & H & 21200 & 7.91 & -6.80 $\pm$ 0.30 & -4.10 $\pm$ 0.30& -4.75 $\pm$ 0.20 & -4.50 $\pm$ 0.30& \citet{Gaensicke2012} \\
PG 0843+517\tablenotemark{o} & H & 24670 & 7.93 & ... & ... &  -4.59 $\pm$ 0.12 & -3.84 $\pm$ 0.18 & {\it this paper}\\
PG 0843+517\tablenotemark{uv} & H & 23095 & 8.17 & -7.30 $\pm$ 0.30 & -5.00 $\pm$ 0.30 & -5.20 $\pm$ 0.20 & -4.60 $\pm$ 0.20 & \citet{Gaensicke2012} \\
\\
GD 40\tablenotemark{o} & He & 15300 & 8.00 & ... & -5.61 $\pm$ 0.09 & -6.76 $\pm$ 0.08 & -6.48 $\pm$ 0.12 & \citet{Klein2010}\\
GD 40\tablenotemark{uv} & He &  15300 & 8.00 &-7.80 $\pm$ 0.20 & -5.68 $\pm$ 0.10 & -6.28 $\pm$ 0.10 & -6.46 $\pm$ 0.10 & \citet{Jura2012}\\
G 241-6\tablenotemark{o} & He & 15300 & 8.00 & ... & -5.60 $\pm$ 0.10 & -6.78 $\pm$ 0.06 & -6.76 $\pm$ 0.06 & \citet{Klein2011} \\
G 241-6\tablenotemark{uv} & He & 15300 & 8.00 & $<$ -8.50 & ... & -6.55 $\pm$ 0.10 & -6.86 $\pm$ 0.10 & \citet{Jura2012}\\
GD 61\tablenotemark{o} & He & 17280 & 8.20 & $<$ -8.80 & ... & -6.85 $\pm$ 0.09 & $<$ -7.50 & \citet{Farihi2011a} \\
GD 61\tablenotemark{uv} & He & 17280 & 8.20 & $<$ -9.10 & -6.00 $\pm$ 0.15 & -6.82 $\pm$ 0.12 & -7.6 $\pm$ 0.3 & \citet{Farihi2013} \\
Ton 345\tablenotemark{o} & He &  18700 & 8.00 & -4.63 $\pm$ 0.19 & -4.58 $\pm$ 0.10 & -4.91 $\pm$ 0.12 & -5.07 $\pm$ 0.10 & \citet{Jura2015}\\
Ton 345\tablenotemark{uv} & He &  19780 & 8.18 & -4.90 $\pm$ 0.20 & -4.25 $\pm$ 0.20 & -4.80 $\pm$ 0.30 & -4.60 $\pm$ 0.20 & \citet{Wilson2015}\\
\enddata
\tablenotetext{o}{ Abundance measurements from the optical data.}
\tablenotetext{uv}{ Abundance measurements from the UV data.}
\end{deluxetable*}

The optical and UV data are often analyzed in different studies and the adopted WD parameters are different, which could be the cause of the discrepancy. We made some preliminary tests and found that we can reproduce the reported abundances using the reported WD parameters. This suggests that this problem is likely to be present in all the WD models used for abundance analysis. Fortunately, the abundance ratios vary much less. For one of the most extreme cases, log n(Fe)/n(Si) is 0.5 in PG~0843+517 in the optical while this number goes up to 0.9 in the UV.

We acknowledge that this presents a challenge in using polluted WDs to derive the chemical compositions of extrasolar planetesimals. This problem only starts to appear when an element is detected both in the optical and UV. There are only eight such systems now and the results are inconclusive. More observations and analysis are needed to understand the extent of this discrepancy. \citet{Gaensicke2012} discussed three possibilities for this discrepancy, i.e. uncertain atomic data, abundance stratification, and genuine variation. They concluded that abundance stratification is the most likely cause. Indeed, this problem is likely related to the atmospheric structure calculation, because optical and UV lines originate from different depths of the WD atmosphere. Solving the optical and UV discrepancy is beyond the scope of current study and will be explored in future work.

\subsection{Circumstellar Gas}
WD~1226+110 and WD~1341+036 have been reported to display calcium infrared triplet emission from circumstellar gas material \citep{Gaensicke2006, Li2017}. Such gas has been detected around 20\% of dusty WDs and it is a result of collision and evaporation of planetesimals. The emission lines can be variable on time-scales of a few years \citep{Wilson2014,Manser2016,Dennihy2018}.

\begin{figure}
\includegraphics[width=\linewidth]{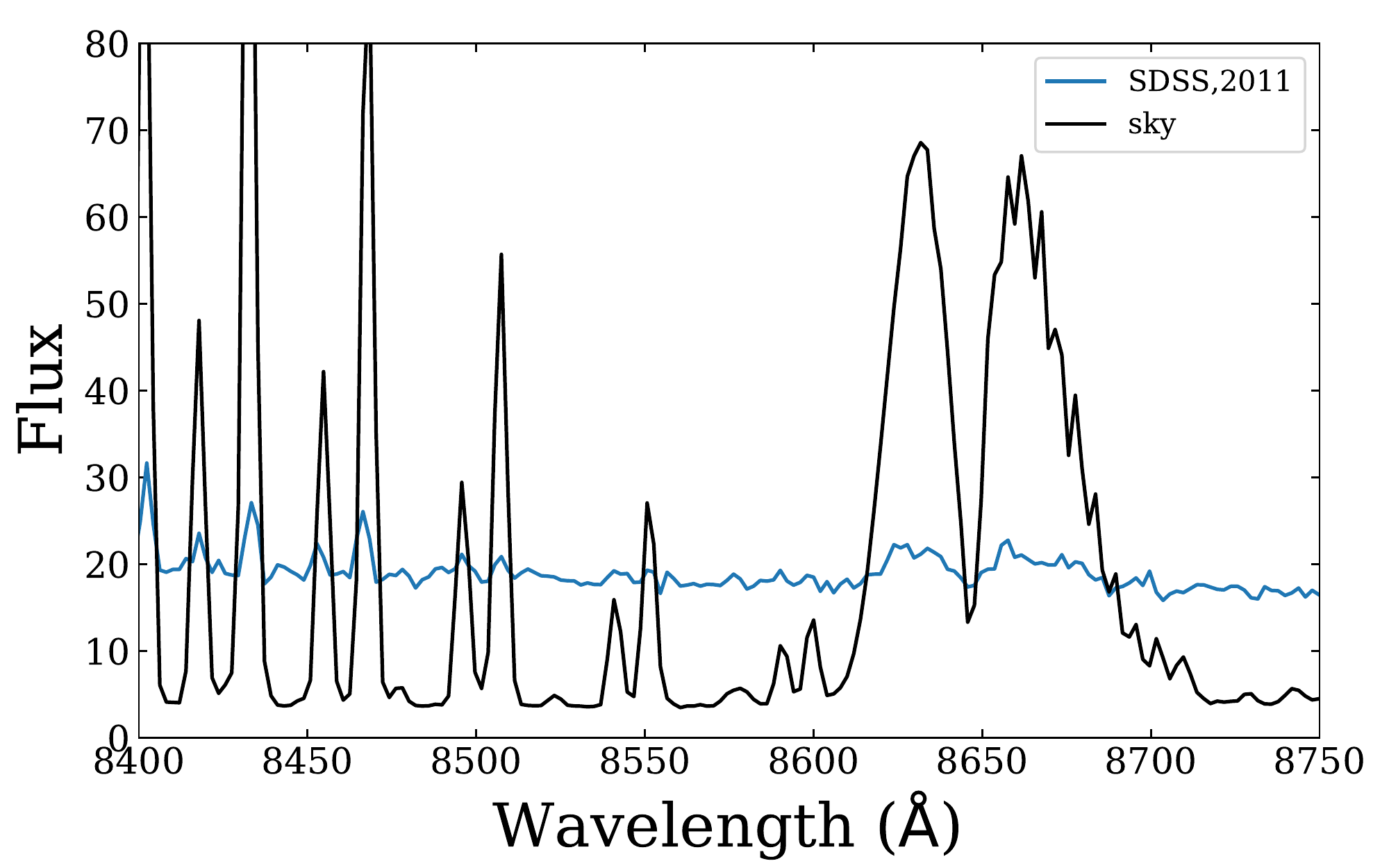}
\caption{SDSS spectra of WD~1341+036 (ID: 4786-55651-0056). The strongest calcium emission at 8662~{\AA} was marginally detected. However, that wavelength also has strong sky lines and the emission is likely due to improper sky subtraction. The calcium triplet emission line was not detected in our ESI data.
\label{fig:gas}}
\end{figure}

For WD~1341+036, no calcium infrared triplet emission was detected in the ESI data. A closer inspection of the original {\it SDSS} discovery spectrum shows that there are strong sky lines at the same wavelengths, as shown in Fig.~\ref{fig:gas}. Likely, the weak emission was due to improper sky subtraction rather than the presence of circumstellar gas. For WD~1226+110, the HIRES data covering the calcium triplet emission feature have been discussed in \citet{Melis2010} and we will not repeat the analysis here. No other WDs analyzed in this sample display calcium triplet emission lines.

\section{Discussion \label{sec:discussion}}

\subsection{Overall Abundance Pattern \label{sec:overall}}

We have detected 8, 9, and 10 heavy elements from the optical spectra of WD~1232+563, WD~1551+171, and WD~2207+121, respectively. Together with 18 other polluted WDs, there now are 21 WDs in total where all the major rock forming elements (i.e. Mg, Si, Ca, Fe, and O) are detected. The 18 systems are G29-38, WD~J0738+1835, HS~2253+8023, G241-6, GD~40, GD~61, PG~1015+161, WD~1226+110, WD~1929+012, PG~0843+517, Ton~345, SDSS~1242+5226, WD~1425+540, WD~1145+017, SDSS~J1043+0855, WD~1536+520, WD~0446-255 and WD~1350-162 \citep{Klein2011, Dufour2012, Gaensicke2012, Jura2012, Wilson2015, Farihi2013, Xu2013a, Jura2015, Raddi2015, Farihi2016, MelisDufour2017, Xu2017, Swan2019}. There are 14 DBZs and except for WD~1425+540, all five major elements are measured in the optical. While for DAZs, aside from WD 1929+012 \citep{Vennes2010}, ultraviolet (UV) spectroscopy is needed to measure all the major elements. Dusty DBZs are the best candidates for detecting multiple heavy elements from the ground.

\begin{figure*}
\includegraphics[width=\textwidth]{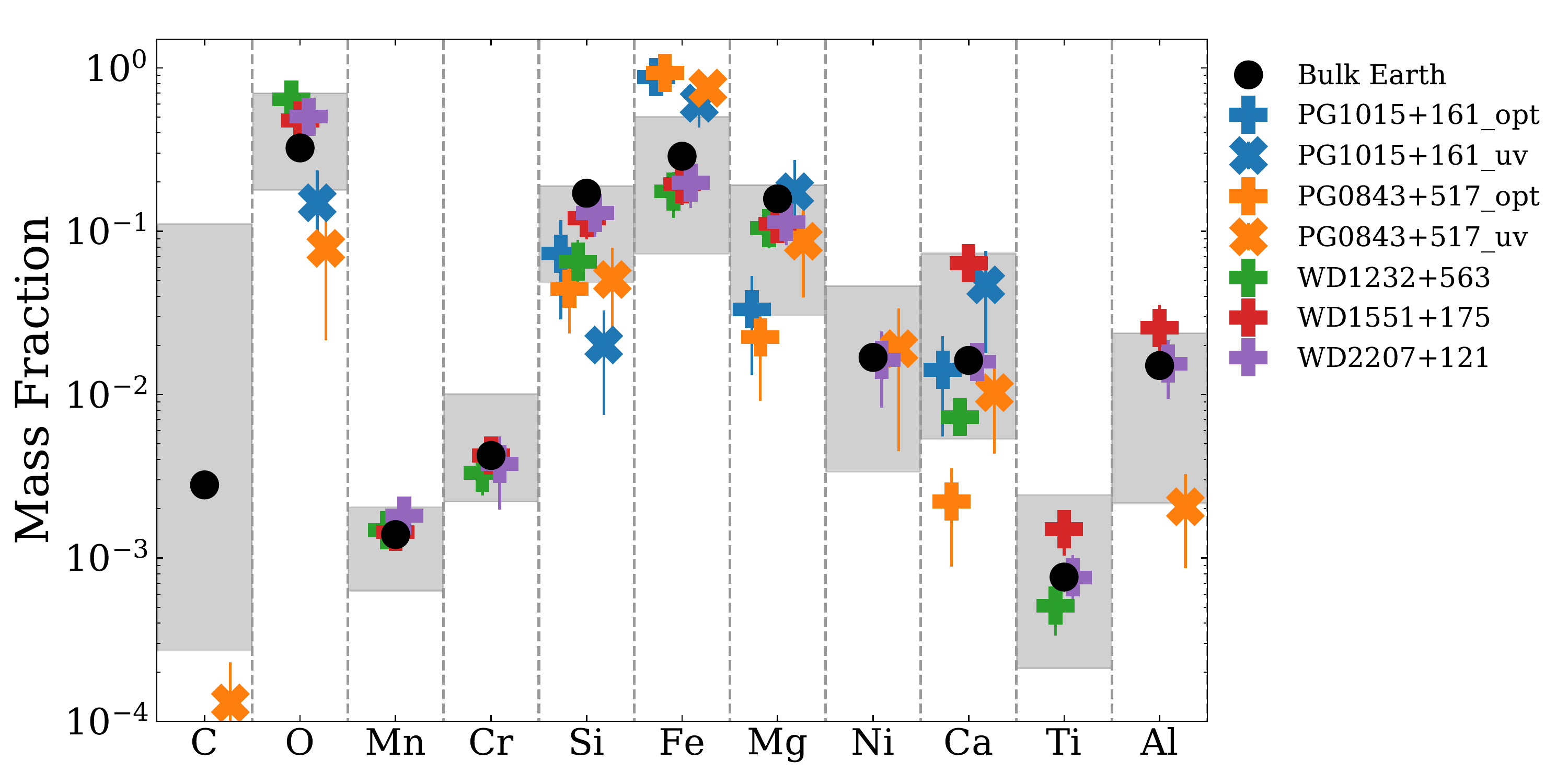}
\caption{Mass fractions of different elements in the accreting material onto 5 WDs analyzed in this paper. Both optical and UV abundances are shown for PG~1015+161 and PG~0843+517. For clarity, only positive detections are marked. The grey shaded area indicates the range observed in all other 16 WDs with the major rock forming elements detected. The elements are arranged in the order of increasing condensation temperature.
\label{fig:comp_DBZ}}
\end{figure*}

For this sample of dusty WDs and candidates, the accretion is likely to be on-going and the WD's atmosphere is possibly dominated by one large parent body. Assuming a steady state accretion, we can calculate the mass fraction for each element, as shown in Fig.~\ref{fig:comp_DBZ}. Even though the exact proportions are different, O, Fe, Si and Mg make up at least 87\% of the total mass for 19 out of 21 WDs, very similar to the composition of bulk Earth. In two systems, carbon comprises a significant fraction of the total mass -- 11\% for WD~1425+540 and 15\% for Ton~345 from the optical and 2.6\% from the UV (see discussion in Section~\ref{sec:discrep}). WD~1425+540 has accreted a significant amount of O, C, and N as well and it has been proposed that the parent body was a Kuiper-Belt-Object analog \citep{Xu2017}. In contrast, Ton~345 has barely accreted enough O to form O-bearing minerals in both optical and UV studies. It has been suggested that the parent body was similar to anhydrous interplanetary dust particles (IDP; \citealt{Jura2015}).

Now we comment on some WDs analyzed in this study.

{\it WD~1232+563: accretion from an O-rich \& H-poor parent body}. Oxygen is 64\% of the total material by mass. Indeed, there is more oxygen than is needed to form MgO, SiO$_2$, Fe$_2$O$_3$\footnote{FeO is a more common form of iron oxide but here we took Fe$_2$O$_3$ to be more conservative with the oxygen budget in the system.}, CaO, and Al$_2$O$_3$; 57\% of the O is uncombined following procedures outlined in \citet{Klein2010}. As a result, H$_2$O is the next most likely carrier of oxygen. There are three other DBs (i.e. GD~61, SDSS~J1242+5226, and WD~1425+540, from \citealt{Farihi2013,Raddi2015,Xu2017} respectively) with a high oxygen content  interpreted as accretion of water-bearing debris. Those three DBs also have a high hydrogen abundance. Hydrogen could either be primordial \citep{KoesterKepler2015} or from accretion over multiple events throughout the WD's cooling history, as H never sinks. Recently, it has been found that there is a correlation between the presence of hydrogen and heavy elements \citep{Gentile-Fusillo2017}. DBs with a large amount of hydrogen may imply accretion of water-rich planetesimals. However, WD~1232+563 does not have enough hydrogen -- only 9\% of the oxygen could be combined with hydrogen to form H$_2$O. This is the only system known so far that has accreted O-rich but H-poor material. The next most likely O carrier is C to form CO$_2$. Unfortunately, from optical observations, we can only derive an upper limit, log n(C)/n(He) $<$ -4.0. The accreting material could be rich in C but still escape optical C detections. 

{\it WD~1551+171: accretion from a refractory-enhanced parent body}: The mass fraction of Ca is 6.4\%, about a factor of 4 higher than the value in bulk Earth. In fact, the mass fractions of all the refractory elements, including Ca, Ti, and Al, are enhanced in WD~1551+171. This object has accreted one of the most refractory rich objects in all WDs, as shown in Fig.~\ref{fig:comp_DBZ}. Likely, the parent body was originally formed at a distance closer to the central star to incorporate a higher fraction of refractory elements, as has been proposed for the formation of refractory-dominated planetesimals \citep{Carter-Bond2012}. However, the accreted material is still not as refractory rich as the calcium aluminum inclusions (CAIs), whose calcium mass fractions can be up to 25\% \citep{Grossman1980}. We found no evidence for refractory dominated planetesimals in our current sample, confirming results from previous studies \citep{JuraXu2013}.

{\it WD~2207+121: accretion from a ``normal" parent body}. Out of the three DBZs analyzed in this study, WD~2207+121 has accreted from a planetesimal that is most similar to the bulk Earth. O, Mg, Si and Fe consist of 95\% of the total mass though there is a slight enhancement of oxygen. To fully assess the nature of the accreting material, the abundances of volatile elements, such as C, S and N, need to be determined. Unfortunately, these elements only have strong transitions in the UV and optical observations are not very constraining \citep[e.g.][]{Jura2012,Gaensicke2012}. Future observations in the UV will better constrain the nature of the accreting object.

{\it PG~1015+161 \& PG~0843+517: accretion from iron-rich objects}. PG~1015+161 has accreted from a parent body with 61\% Fe by mass from UV studies and the fraction goes up to 88\% using optical abundances. For PG~0843+517, it is 75\% from the UV and 93\% from the optical, as shown in Fig.~\ref{fig:comp_DBZ}. Even though there are discrepancies between the absolute abundances, the relative Fe fractions are very high and these are among the most iron-rich objects known to date. Possibly, they are a result of accretion of core-like material \citep{Harrison2018}. Previously, accretion from iron-rich objects has been identified in two other DZs, i.e. SDSS~J0823+0546 and SDSS~J0741+3146 \citep{Hollands2018}. A potential issue there is that neither Si nor O is detected in those two systems and it is hard to derive the nature of the material without the abundances of all the major elements. 

\subsection{Dusty v.s. non-Dusty WDs}

\begin{figure*}
\gridline{\fig{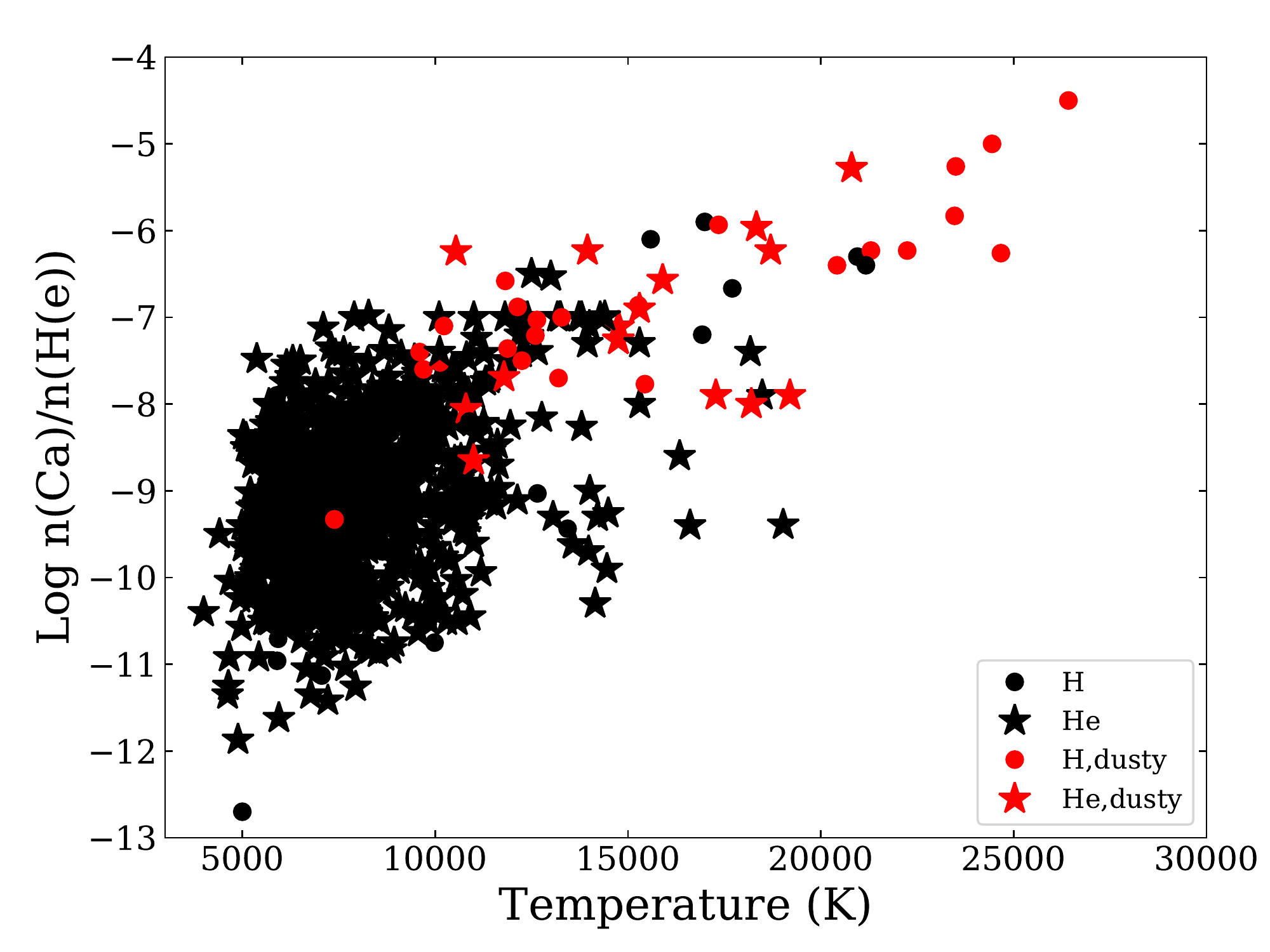}{0.5\textwidth}{}
\fig{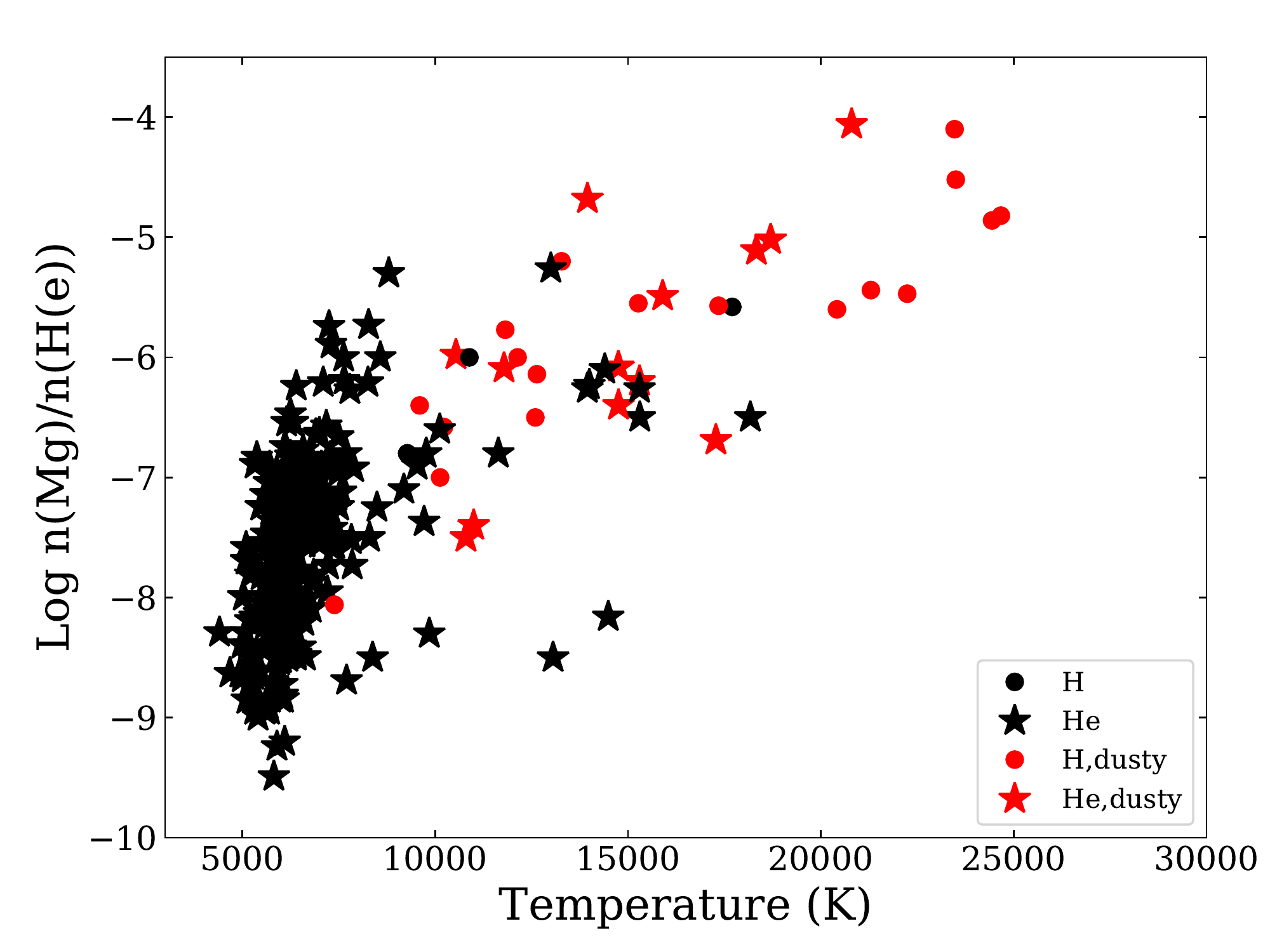}{0.5\textwidth}{} }
\gridline{\fig{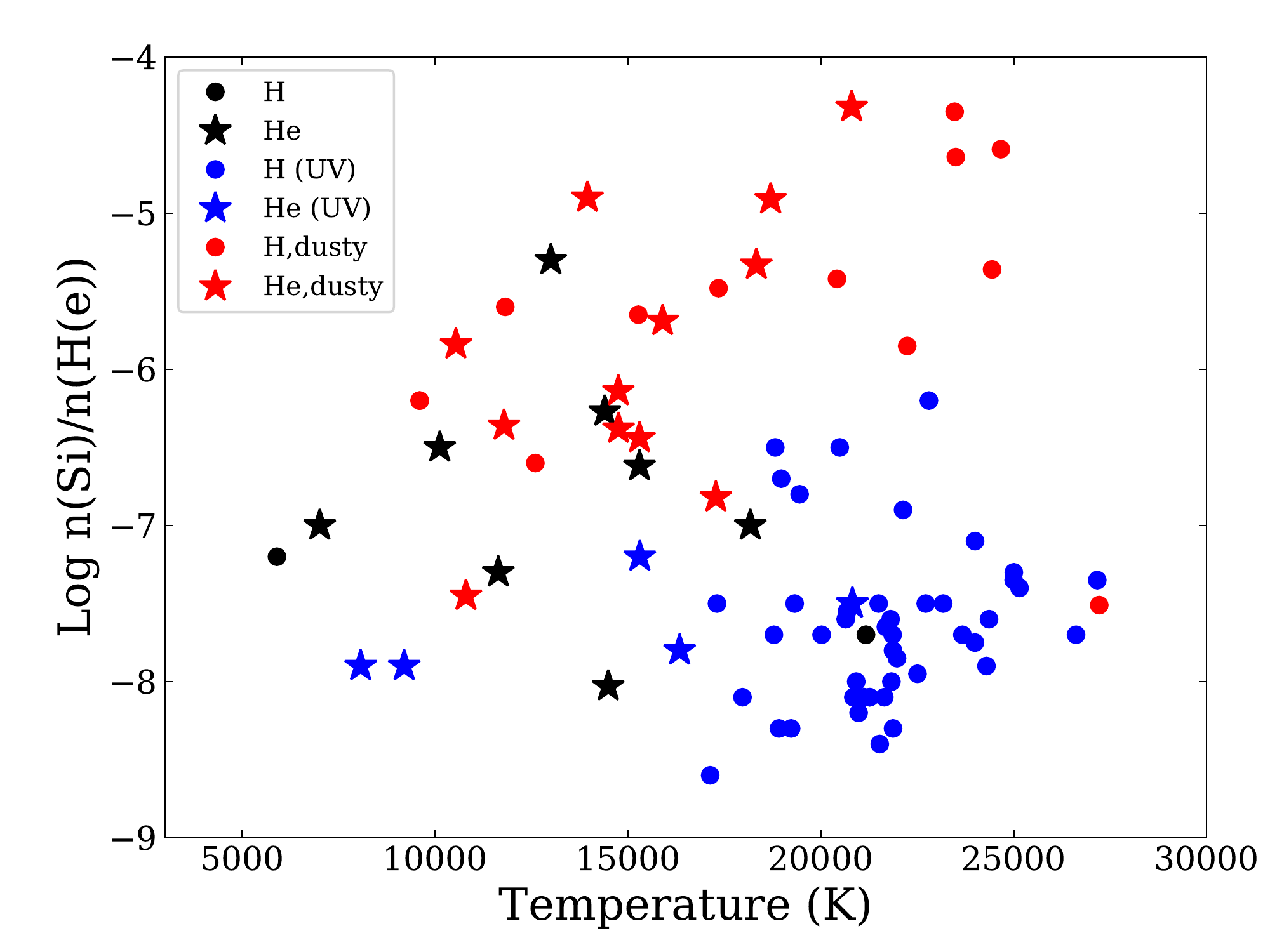}{0.5\textwidth}{}
\fig{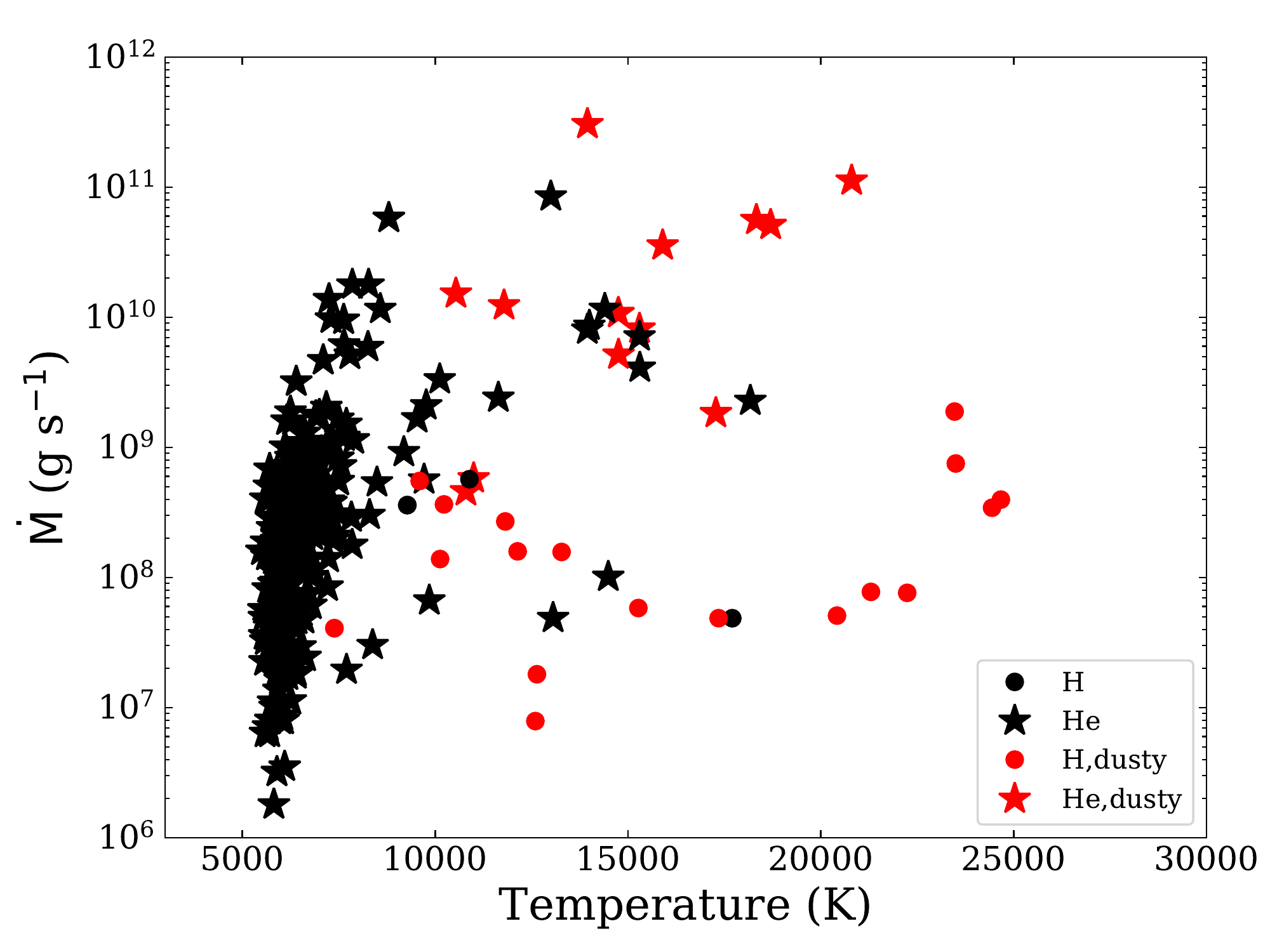}{0.5\textwidth}{} }
\caption{Ca, Mg, Si abundances, and the overall accretion rates in polluted WDs as a function of WD temperature. All the abundances are from optical observations, except for the blue points, which are from the UV.
\label{fig:Z-T}}
\end{figure*}

Between 1-4\% of WDs display infrared excess from dust disks, but the fraction of WDs that show pollution can be as high as 50\% \citep{Zuckerman2003, Zuckerman2010, Barber2014,Koester2014a,Wilson2019}. Here, we compare pollution levels in WDs with and without a dust disk, as shown in Fig.~\ref{fig:Z-T}. In addition to WDs analyzed in this study, we also include the 16 WDs with all the major elements detected (see Section~\ref{sec:overall}) and some WD abundances from the literature \citep{Koester2011,Koester2014a,KoesterKepler2015, Hollands2018}. The total mass accretion rate is calculated from the magnesium abundance, assuming that it is 15.8\% of the total mass, as that in bulk Earth \citep{Allegre2001}. WDs cooler than 5000~K have no diffusion timescales calculated from the Montreal White Dwarf Database \citep{Dufour2017} and are excluded in this figure. Previous work often extrapolates the overall accretion rate from calcium, which is the most easily detectable heavy element in the optical. However, as shown in Fig.~\ref{fig:comp_DBZ}, the mass fraction of calcium varies by two orders of magnitude in different WDs while the mass fraction of magnesium has a much smaller spread. In addition, magnesium is a major element in the accreting material and we consider the numbers derived from magnesium a better representation of the true accretion rate.

\begin{deluxetable*}{lccccccccc}
\tablecaption{Pollution Level Comparison \label{tab:comp_Z}}
\tablewidth{0pt}
\tablehead{
\colhead{} &\colhead{H} & \colhead{H, dusty} & \colhead{He} & \colhead{He, dusty}
}
\startdata
Log n(Ca)/n(H(e)) & -7.11 $\pm$ 1.40 & -5.64 $\pm$ 1.25 & -8.28 $\pm$ 1.44 & -6.24 $\pm$ 0.96 \\
Log n(Mg)/n(H(e)) & -6.02 $\pm$ 0.54 & -5.10 $\pm$ 0.95 & -6.94 $\pm$ 1.49 & -4.99 $\pm$ 0.98 \\
Log n(Si)/n(H(e)) & -7.24 $\pm$ 0.83\tablenotemark{a} & -5.06 $\pm$ 0.65 & -6.33 $\pm$ 1.18 & -5.15 $\pm$ 0.82\\
Log $\dot{\mathrm{M}}$ (g s$^{-1}$) & 8.68 $\pm$ 0.22 & 8.48 $\pm$ 0.63 & 9.13 $\pm$ 1.99 & 10.68 $\pm$ 0.74 \\
\enddata
\tablenotetext{a}{This is mostly from UV measurements in \citet{Koester2014a}.
}
\tablecomments{These are average values for the WDs presented in Fig.~\ref{fig:Z-T}.}
\end{deluxetable*}

Before comparing properties of different systems, there are three systematics to be aware of:

{\it Detection limits of heavy elements}: it is much harder to detect heavy elements in hot WDs compared to cool ones. For example, the equivalent width of Ca II 3933~{\AA} is 30m{\AA} in PG~0843+517. At the temperature of 21700~K, a calcium abundance log n(Ca)/n(H) = -6.26 is derived (see Table~\ref{tab:DAZ}). In comparison, for WD 1344+106, a DA at 6945 K with a similar strength of Ca II 3933, the calcium abundance is -11.13 \citep{Zuckerman2003}.

{\it Detection limits of dust disks}: It is still an open question why some polluted WDs have a disk while others do not. It could be that there are small disks that escaped infrared detection or the accretion is supplied by a pure gas disk that does not produce any spectroscopic signatures; alternatively, the dust disk can be completely accreted but the heavy elements have not fully settled \citep{Jura2008,Bonsor2017}. As shown in Fig.~\ref{fig:Z-T}, most dust disks are detected around hot WDs and there are very few around WDs cooler than 10,000~K \citep{XuJura2012}. As a result, the dusty and non-dusty WDs shown in Fig.~\ref{fig:Z-T} span different temperature ranges.

{\it Optical \& UV discrepancy}. Most of the abundances are from optical observations, except for Si in hot DAs (blue dots in Fig.~\ref{fig:Z-T}). These are from {\it HST} UV observations reported in \citet{Koester2014a}. Admittedly, UV spectroscopy is more sensitive to small amounts of pollution. But it is still surprising that none of the DAs in the UV sample have as high of a Si abundance as those DAs observed in the optical. This presents an additional support of a systematic optical and UV discrepancy discussed in Section~\ref{sec:discrep}

With these three caveats in mind, here is a summary of the findings based on Fig.~\ref{fig:Z-T} and Table~\ref{tab:comp_Z}.

{\it The accretion rates in He-dominated WDs are higher than those in H-dominated WDs.} The largest difference is between dusty DAZs and DBZs, whose average accretion rates differ by a factor of 100. DBs have much longer settling times ($\sim$ 10$^5$ yr) and the accretion rate represents an average historical rate. While in DAs, the settling times are $\sim$ 1 to 10$^3$ yr and the accretion rate is essentially instantaneous. As a result, DAs probe continuous events while DBs probe long-term activity \citep{Farihi2012b}. The difference in the accretion rate can be used to constrain the accretion history: small planetesimals are accreted continuously while accretion of larger bodies are stochastic events \citep{Wyatt2014}. Alternatively, the difference in the accretion rates could be related to inaccurate WD modeling, e.g. convective overshooting, thermohaline effect, etc \citep[e.g.][]{BauerBildsten2019, Cunningham2019}.

{\it There is an overall trend of decreasing pollution level as a WD cools. But this trend disappears after correcting for element settling and the change of the convection zone size.} The absolute Ca and Mg abundances decrease as the WD cools and this can be explained as a result of a decreasing amount of unstable minor planets around cooler WDs \citep[e.g.][]{Chen2019}. However, we find no significant trend for the overall accretion rate, which is related to absolute abundances, convection zone size, and settling time, as a function of the WD temperature. A statistical analysis using the calcium abundances arrives at the same conclusion -- there is no age dependence of the accretion rate \citep{Wyatt2014}. This constant accretion rate poses a challenge for dynamical studies to continuously produce tidal disruption events over billions of years \citep[e.g.][]{FrewenHansen2014}.

{\it There is no strong difference between WDs with and without a dust disk, in terms of absolute calcium, magnesium, silicon abundances, and the overall mass accretion rate.} The high pollution level and dust disk connection has been recognized since \citet{vonHippel2007} and confirmed in follow up studies \citep[e.g.][]{Bergfors2014}. However, we caution here that the spread is big and pollution is only slightly elevated in dusty WDs. In addition, there is not much overlap in the temperature range between dusty and non-dusty WDs. For non-dusty WDs, the calcium and magnesium measurements are mostly around WDs cooler than 10,000 K while dusty WDs are all between 10,000 K and 25,000~K. Due to the detection limit, only the more heavily polluted hot WDs can be identified. Future observations around hot non-dusty WDs can help filling in the gap and understanding the potential differences between dusty and non-dusty WDs.

\subsection{Mg/Ca - Si/Ca Ratios \label{sec:ratio}}

\begin{figure}
\includegraphics[width=0.5\textwidth]{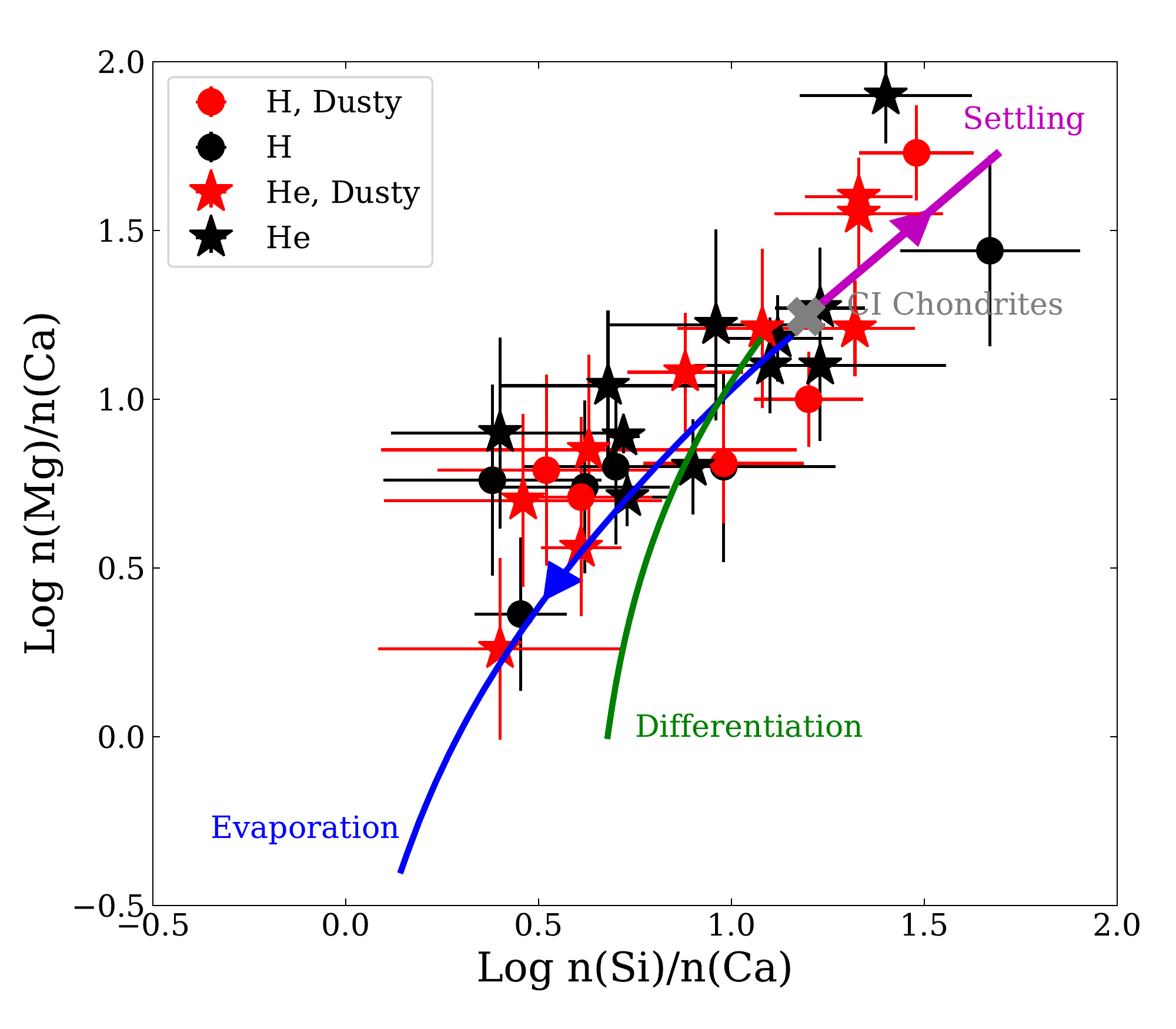}
\caption{Mg/Ca - Si/Ca ratios in polluted WDs. There is a large spread in these ratios but there is not any clear difference between DAs and DBs, or between WDs with and without a dust disk. Starting with the composition of CI chondrites (grey cross), we computed three models, i.e. differential settling (magenta), igneous differentiation (green), and evaporation (blue). The spread of these ratios can be explained as a result of evaporation. Alternatively, the spread can be from the intrinsic difference of material that pollutes the WDs. 
\label{fig:evaporation}}
\end{figure}

In the optical, Ca, Mg, and Si are the most commonly detected elements in polluted WDs. A comparison of their abundance ratios is shown in Fig.~\ref{fig:evaporation}. The Mg/Ca ratios vary from 1 to 60 while the Si/Ca ratios vary from 1 to 45; the spread is much larger than that which is measured in CI chondrites. 

No difference is found between the abundance ratios measured in DAZs and DBZs. This is not surprising because we did not expect any differences between compositions of planetary debris around DAs and DBs. However, DAZs tend to have short settling times, e.g. days to years, much shorter than 10,000 -- 1000,000 yrs found for DBZs. Fig.~\ref{fig:evaporation} shows that settling does not play a significant role in the abundances measured in polluted WDs.

Somewhat surprisingly, Fig.~\ref{fig:evaporation} does not show any difference between the abundance ratios for WDs with and without a dust disk either. Previously, it has been suggested that WDs without a disk might be accreting from many smaller planetesimals while WDs with an infrared excess are accreting from one large planetesimal \citep{Jura2008}. A mix of planetesimals could create different chemical signatures in polluted WDs but it is not observed in this sample.

The spread in the measured Mg/Ca and Si/Ca ratios could come from the intrinsic difference of the material polluting WDs. Here, we explore three additional processes that could lead to the spread.

\subsubsection{Differential Settling \label{subsec:settling}}

The accretion/diffusion scenario in a WD's atmosphere has been studied extensively and there are three main stages, i.e. build-up, steady-state, and decaying. \citep{Dupuis1993a, Koester2009a}. In the build-up stage, the measured abundance ratios in the WD's photosphere is equal to the abundance ratios in the parent body.

\begin{equation}
\frac{n(A)}{n(B)}_\mathrm{WD} = \frac{n(A)}{n(B)}_\mathrm{par}
\label{equ:buildup}
\end{equation}

In the steady-state, the measured abundance ratios are modified by the ratio of settling times, which is denoted as $\tau_\mathrm{A}$ and $\tau_\mathrm{B}$, respectively. In this case,

\begin{equation}
\frac{n(A)}{n(B)}_\mathrm{WD} = \frac{n(A)}{n(B)}_\mathrm{par} \times \frac{\tau_\mathrm{A}}{\tau_\mathrm{B}}
\label{equ:steady}
\end{equation}

In the decaying phase, accretion has stopped and the abundances decrease exponentially. The abundance ratios are dependent on the time, t, from which the accretion has stopped. The decaying phase displays the largest variations in terms of abundance ratios.

\begin{equation}
\frac{n(A)}{n(B)}_\mathrm{WD} = \frac{n(A)}{n(B)}_\mathrm{par} \times \frac{e^{-t/\tau_\mathrm{A}}}{e^{-t/\tau_\mathrm{B}}}
\label{equ:decay}
\end{equation}

WDs with a dust disk are often assumed to be in steady-state accretion (Equ.~\ref{equ:steady}) while WDs without a disk can be in the any of the three stages (Equ.~\ref{equ:buildup} - \ref{equ:decay}). 

Starting with the composition of CI chondrites, we explore the effects of differential settling on abundance ratios.   Even though the settling timescales are dependent on WD parameters, for polluted WDs in Fig.~\ref{fig:evaporation} (T$>$10,000 K), Ca always has the shortest settling time compared to Mg and Si. From build-up (Equ~\ref{equ:buildup}) to steady-state (Equ~\ref{equ:steady}) to decaying phase (Equ~\ref{equ:decay}), Si/Ca and Mg/Ca both become larger and the ratio moves to the top right of Fig.~\ref{fig:evaporation}. We experimented with different sets of WD parameters and the results are similar. The settling calculation presented in Fig.~\ref{fig:evaporation} is for a helium-dominated atmosphere at 10,000~K. The changes of element ratios from differential settling cannot explain the observed spread. This is further supported by the fact that the scatter is observed in both DAZs and DBZs, which have orders of magnitude difference in settling times. In summary, differential settling cannot explain the large spread in Si/Ca and Mg/Ca ratios observed in polluted WDs.

\subsubsection{Igneous Differentiation}

The large spread in the Fe fractions in polluted WDs have been interpreted as a result of igneous differentiation and collision \citep[e.g.][]{Xu2013a,Jura2013b}. For example, WDs accreting from fragments of a differentiated body, i.e. crust or core, would display different fractions of Fe. Here, we explore the effect of igneous differentiation on the Mg/Ca-Si/Ca ratios. 

The green curve in Fig.~\ref{fig:evaporation} represents our calculation for igneous differentiation using the pMelts program \citep{Ghiorso2002} with an Earth-like mantle starting composition at low pressure.  The curve shows the melt composition as it evolves towards higher concentrations of Si and Ca and lower concentrations of Mg. There is a continuous removal of crystals as the melt cools from 1800~K to 1470~K.  Even though igneous differentiation would change Si/Ca and Mg/Ca ratios in the same direction as the data, the curve gives a poor overall fit.  

\subsubsection{Evaporation}

Another possibility is that we are witnessing an evaporation sequence of disrupted planetesimals surrounding the WDs. The chemical effects of evaporation are determined by two competing processes, i.e. the rate at which the evaporating fragment shrinks and the rate of element diffusions within the melting fragment itself. If the fragment evaporates much more rapidly than the element diffusion time, the composition of the accreting material is unaffected by evaporation. Conversely, if the rate of element diffusion is faster than the rate of evaporation, an evaporation sequence will be observed in the abundances patterns of the WD's atmosphere.

Here, we model evaporation using the methods described by \cite{Young2019} and \cite{Young1998}. In these calculations, a CI chondrite bulk composition is projected into the CaO-MgO-Al$_2$O$_3$-SiO$_2$ (CMAS) system where the evaporation rates are well calibrated experimentally \citep{Richter2002}. We calculated a representative curve of chemical fractionation from evaporation for a 10$^{13}$g (100m radius) object evaporating at 1500~K, as shown in Fig.~\ref{fig:evaporation}.

\begin{figure}
\centering
\includegraphics[width=0.5\textwidth]{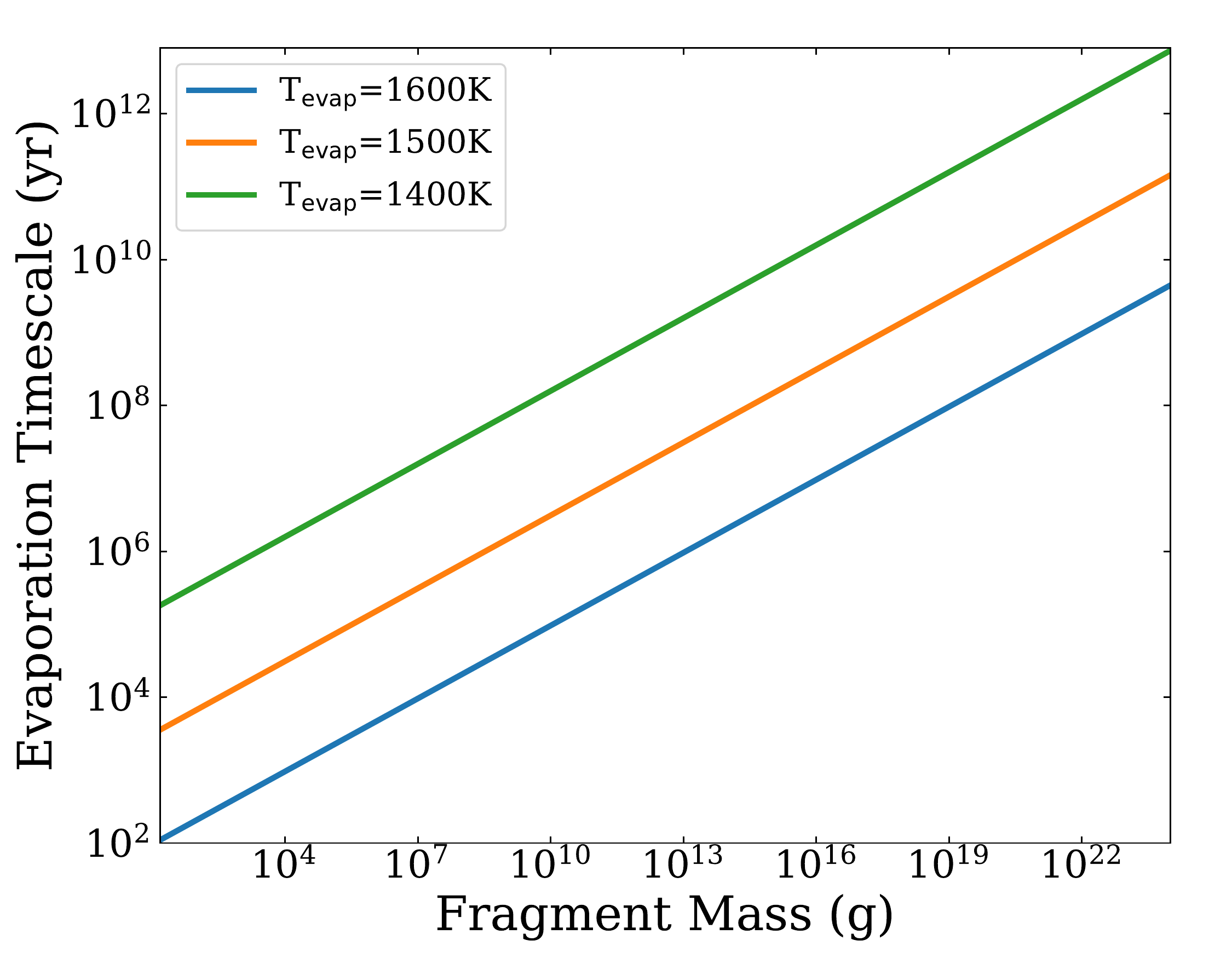}
\caption{Evaporation timescale as a function of fragment mass. Three different evaporation temperatures were considered, T$_\mathrm{evap}$ = 1600 K, 1500 K, and 1400 K, respectively.
\label{fig:app}}
\end{figure}

Based on the similarity of the trajectory of evaporation and the polluted WD data in Fig.~\ref{fig:evaporation}, we speculate that chemical fractionation by evaporation might be an explanation for the spread in Mg/Ca and Si/Ca ratios. Under this scenario, the evaporation timescale must be much longer than both the accretion time and the settling time. The accretion timescales are quite uncertain. One limit is the Poynting-Roberson drag timescale, which is several hundred years for 100~$\mu$m grains around WDs \citep{Rafikov2011a,Bonsor2017}. The longest element settling time is $\sim$~10$^6$~yr for cool DBZs in this sample. Fig.~\ref{fig:app} explores evaporation timescale as a function of fragment mass. For an evaporation timescale longer than 10$^6$ yr, a minimum fragment mass of 10$^8$~g is required for an evaporation temperature of 1500~K. 

\begin{figure}
\centering
\includegraphics[width=0.5\textwidth]{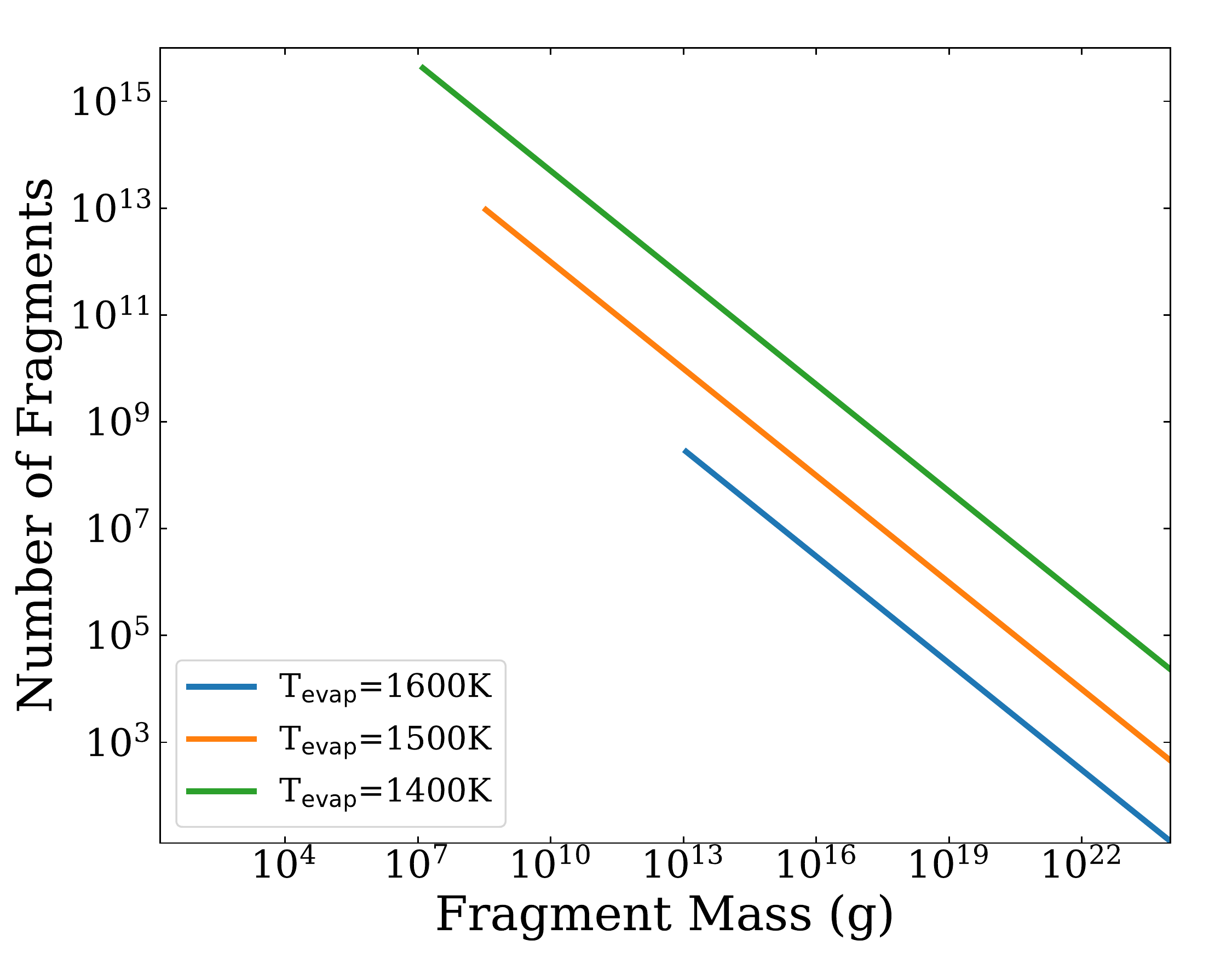}
\caption{Number of fragments as a function of fragment mass in order to produce an accretion rate of 10$^8$ g s$^{-1}$ as observed in many polluted WDs. We include three different evaporation temperatures, i.e.1600 K, 1500 K, and 1400 K.
\label{fig:N-M}}
\end{figure}

An overall mass accretion rate of 10$^8$ g s$^{-1}$ has been observed around many polluted WDs (see Fig.~\ref{fig:Z-T}). To reproduce this accretion rate, we calculated the number of fragments as a function of the fragment mass in Fig.~\ref{fig:N-M}. The lower limit to the fragment mass is determined by the evaporation timescale, which needs to be larger than 10$^6$~yr; the upper limit is bound by the accretion rate of 10$^8$ g s$^{-1}$. Likely, the parent body was broken up into many pieces due to tidal forces and we are witnessing several fragments evaporating at the same time. This process has been observed around WD~1145+017, where transits from at least six stable periods were identified within a small semi-major axis \citep{Vanderburg2015}. These transits are probably all produced from fragments coming from one big parent body. 

\section{Conclusions \label{sec:conclusion}}

In this paper, we report results from spectroscopic studies of 19 WDs with infrared excesses from a dust disk or disk candidates. A majority of the data were taken with Keck/HIRES and Keck/ESI. Two systems (WD~1226+110 and WD~1346+031) in our sample were perviously reported to display calcium infrared triplet emission lines from circumstellar gas. Our observation shows that the gas emission around WD~1346+031 is likely not real. 

The main results are summarized as follows.

$\bullet$ To zeroth order, the chemical compositions of extrasolar planetary material resembles that of bulk Earth, as shown in Fig.~\ref{fig:comp_DBZ}. However, we are starting to have a large enough sample to probe different compositions. WD~1232+563 is accreting from O-rich but H-poor material; WD~1551+171 is accreting from a refractory-rich object while PG~1015+161 and PG~0843+517 are accreting from Fe-rich objects. 

$\bullet$ There is no strong difference in the pollution level between WDs with and without a dust disk (see Fig.~\ref{fig:Z-T}). However, we caution that there are at least three caveats in the analysis, i.e. detection limits of heavy elements, detection limits of dust disks, and optical and UV discrepancies. A uniform sample of polluted WDs is needed to understand the overall pattern between dusty and non-dusty WDs. 

$\bullet$ There is a spread in the Mg/Ca and Si/Ca ratios in polluted WDs that cannot be explained by differential settling or igneous differentiation. Chemical fractionation from evaporation fits the data the best, as illustrated in Fig.~\ref{fig:evaporation}. In this scenario, we can constrain the mass and number of evaporating bodies. Alternatively, it could come from intrinsic variations of the compositions of extrasolar planetary debris.

An outstanding issue in this work is the discrepancy of elemental abundances derived from the optical and UV observations; fortunately, element abundance ratios are not as heavily impacted. Self-consistent WD modeling is required to understanding the origin and push the polluted WD studies to a higher precision to study different planet formation and evolution processes.

\smallskip
{\it Acknowledgements.} We thank A. Bedard and P. Bergeron on useful discussions about deriving WD parameters, E. Dennihy on discussing the {\it SDSS} data of WD~1341+036, D. Koester on modeling PG~0010+280, and A. Bonsor, J. Harrison for useful comments on the compositions of the accreting material. Majority of the data presented herein were obtained at the W. M. Keck Observatory, which is operated as a scientific partnership among the California Institute of Technology, the University of California and the National Aeronautics and Space Administration. The Observatory was made possible by the generous financial support of the W. M. Keck Foundation. The authors wish to recognize and acknowledge the very significant cultural role and reverence that the summit of Maunakea has always had within the indigenous Hawaiian community.  We are most fortunate to have the opportunity to conduct observations from this mountain.This work also uses results from the {\it Sloan Digital Sky Survey} ({\it SDSS}) and the European Space Agency (ESA) space mission {\it Gaia}.

This work is supported by the Gemini Observatory, which is operated by the Association of Universities for Research in Astronomy, Inc., on behalf of the international Gemini partnership of Argentina, Brazil, Canada, Chile, the Republic of Korea, and the United States of America. Additional support for this work was provided by NASA through grant number \#14117 from the Space Telescope Science Institute, which is operated by AURA, Inc., under NASA contract NAS 5-26555. This work is also based on observations collected at the European Organisation for Astronomical Research in the Southern Hemisphere under ESO programme 096.C-0132.

\software{IRAF \citep{Tody1986, Tody1993}, \href{http://www.astro.caltech.edu/~tb/makee/}{MAKEE}, Matplotlib \citep{Matplotlib}}

\appendix

\section{Observing Log \& Figures}

\begin{figure*}
\includegraphics[width=0.83\textwidth]{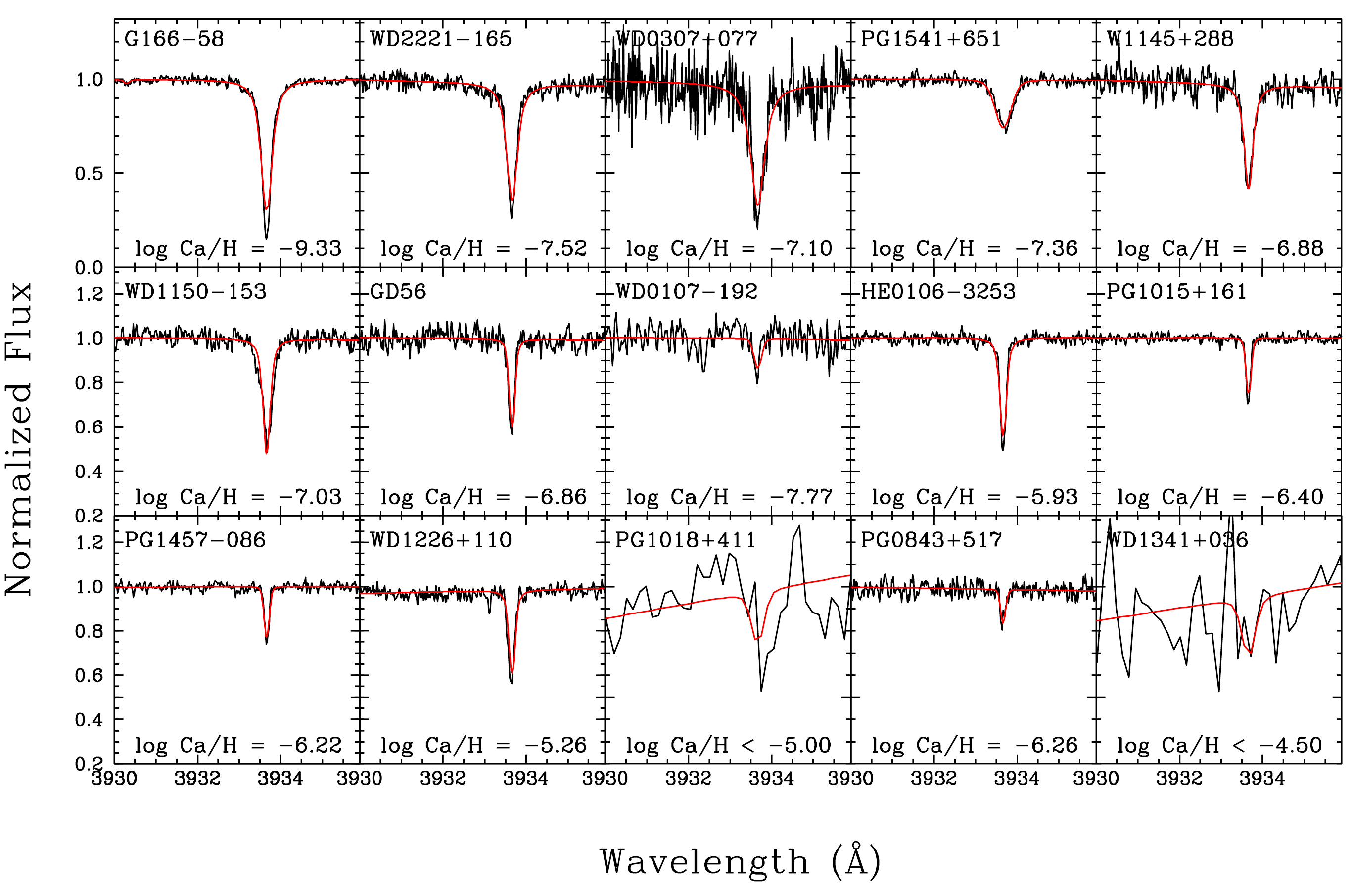}
\caption{Calcium K 3933~{\AA} line region for DAZs. The black and red lines represent the data and our best fit model, respectively. From top left to bottom right, the WDs are arranged in increasing effective temperature.
\label{fig:DAZCa}}
\end{figure*}

\begin{figure*}
\includegraphics[width=0.83\textwidth]{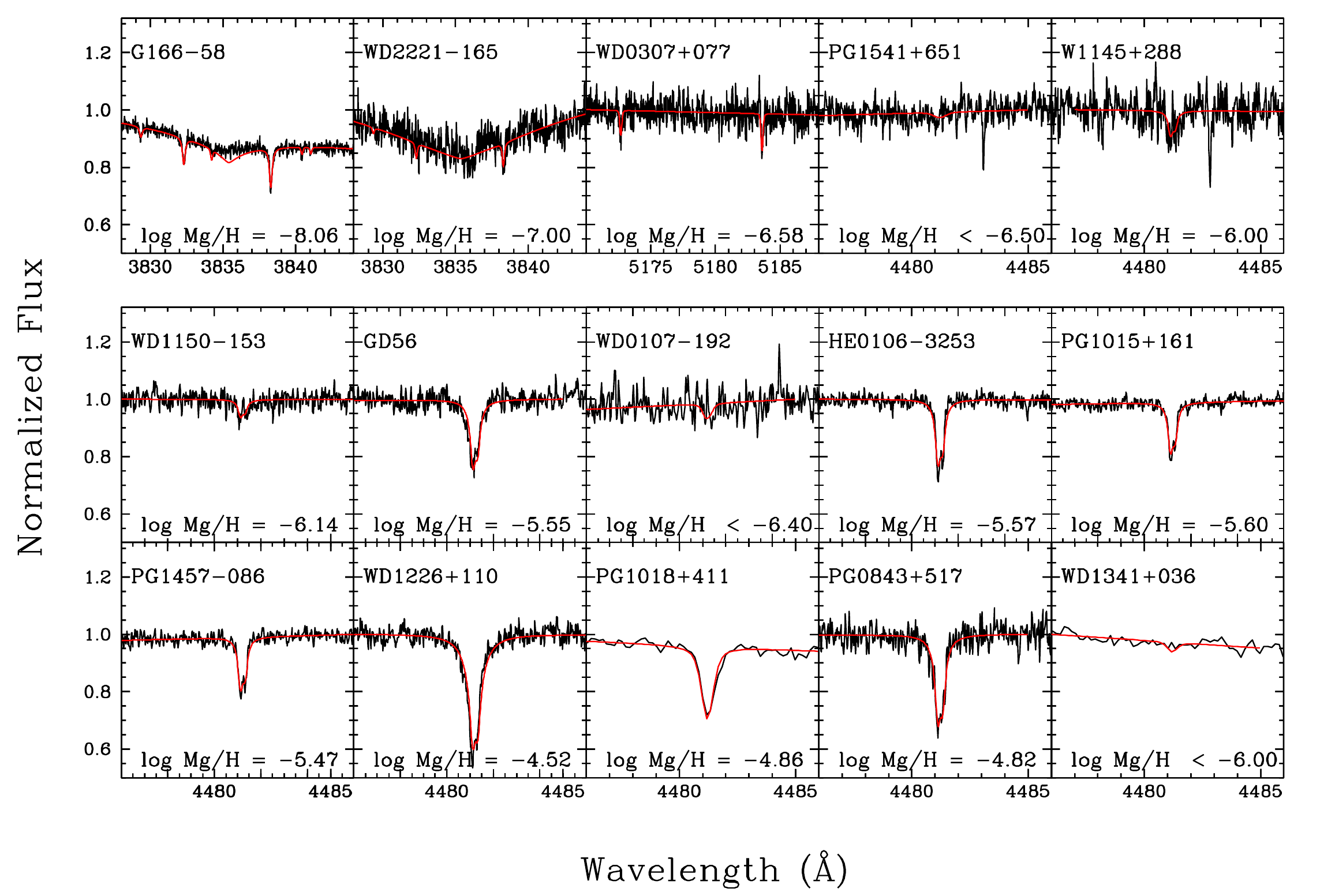}
\caption{Same as Fig.~\ref{fig:DAZCa} except around Mg region. 
\label{fig:DAZMg}}
\end{figure*}

\begin{figure*}
\includegraphics[width=\textwidth]{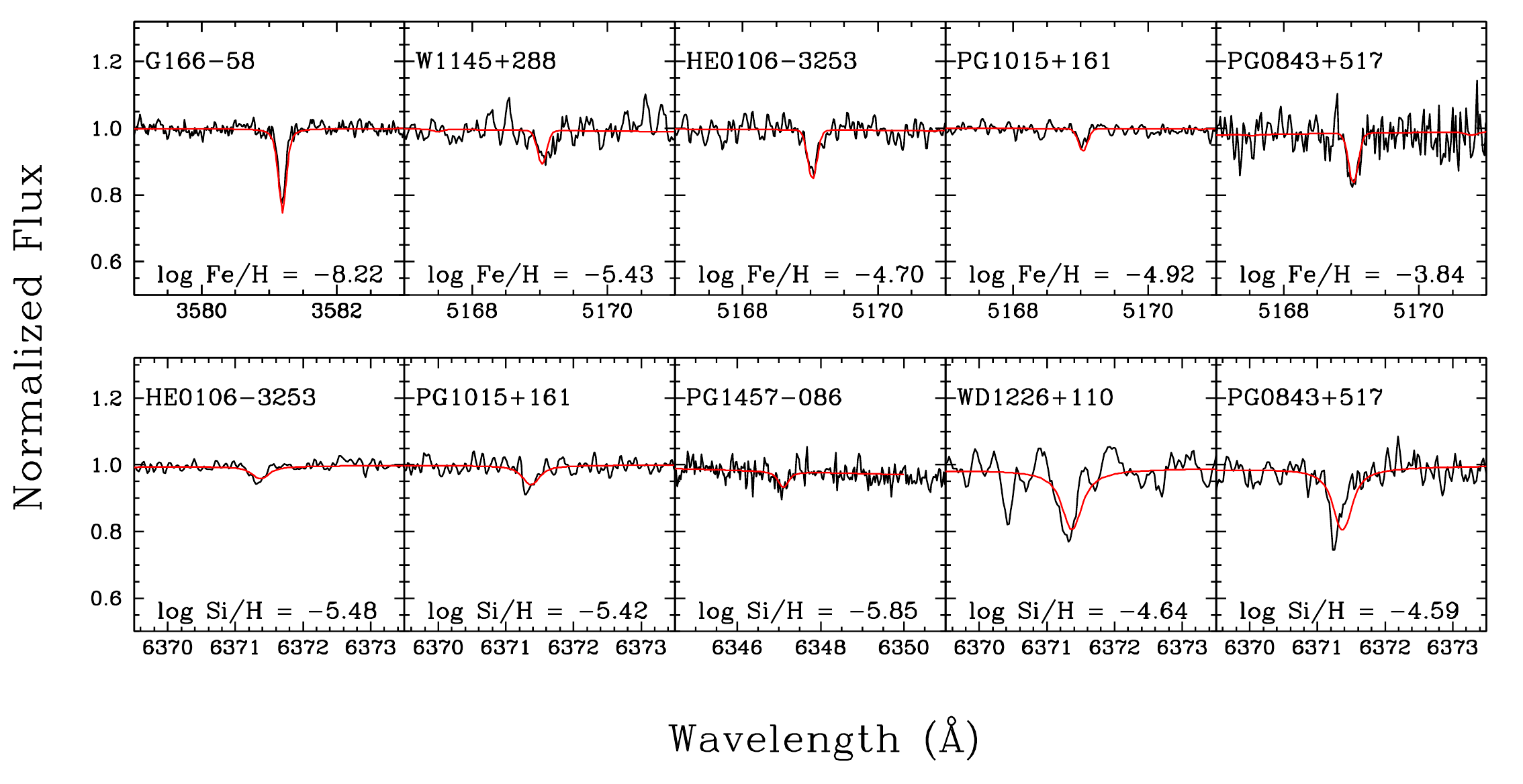}
\caption{Same as Fig.~\ref{fig:DAZCa} except around Fe and Si region.
\label{fig:DAZFeSi}}
\end{figure*}

\begin{figure*}
\includegraphics[width=0.8\textwidth]{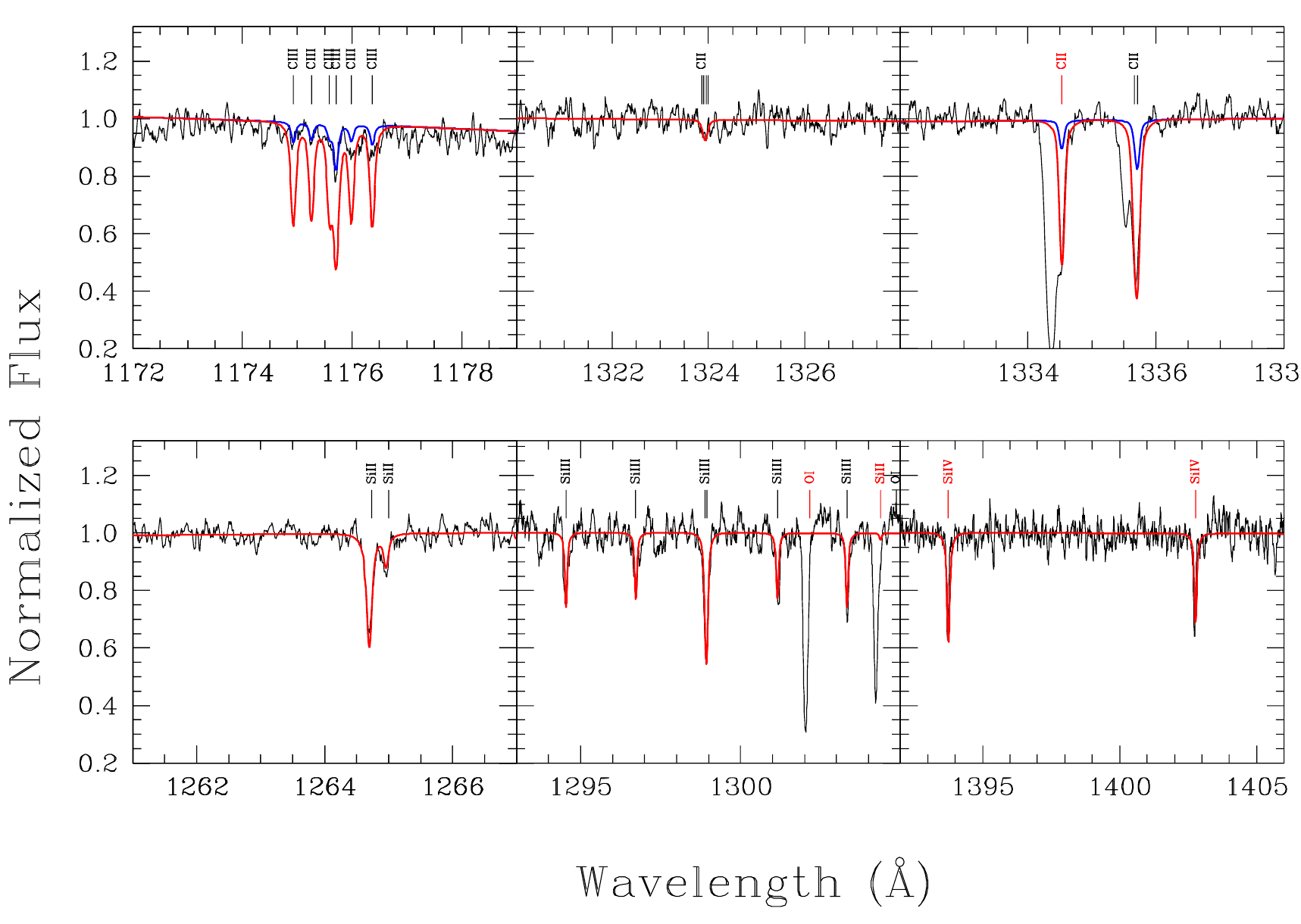}
\caption{{\it HST}/COS data for PG0010+280. For carbon, two sets of models are shown, one with log n(C)/n(H) = -8.01 (blue line) that fits the C III lines; another one with log n(C)/n(H) = -6.75 (red line) that fits C II better. Red labels represent transitions from the ground state; they also show absorption from the interstellar medium in features that are blue-shifted relative to the photospheric component.
}
\label{fig:Fit_PG0010+280}
\end{figure*}

\begin{figure*}
\includegraphics[width=\textwidth]{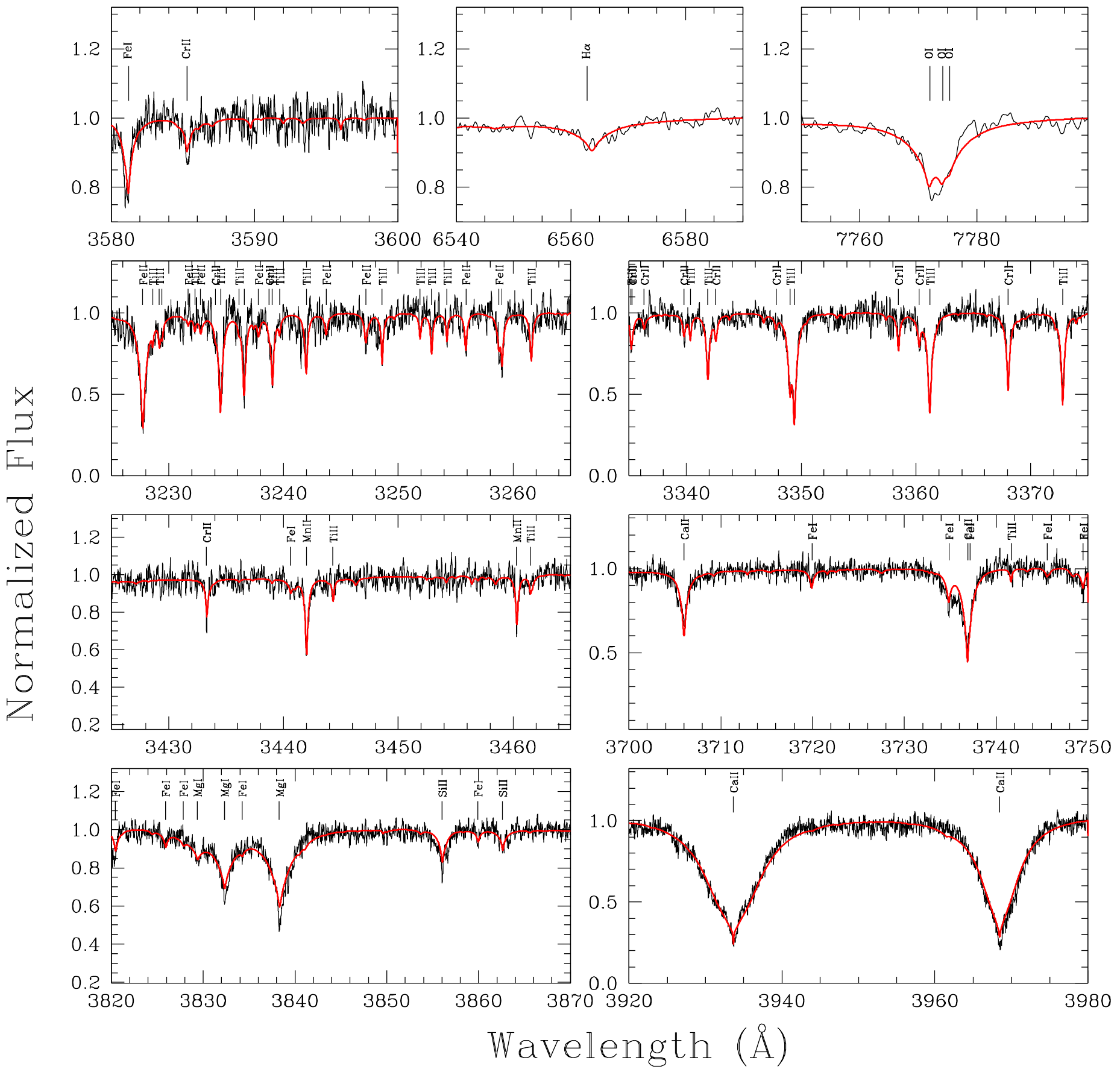}
\caption{Best fit model on the spectra of WD~1232+563.}
\label{fig:Fit_WD1232+563}
\end{figure*}

\begin{figure*}
\includegraphics[width=\textwidth]{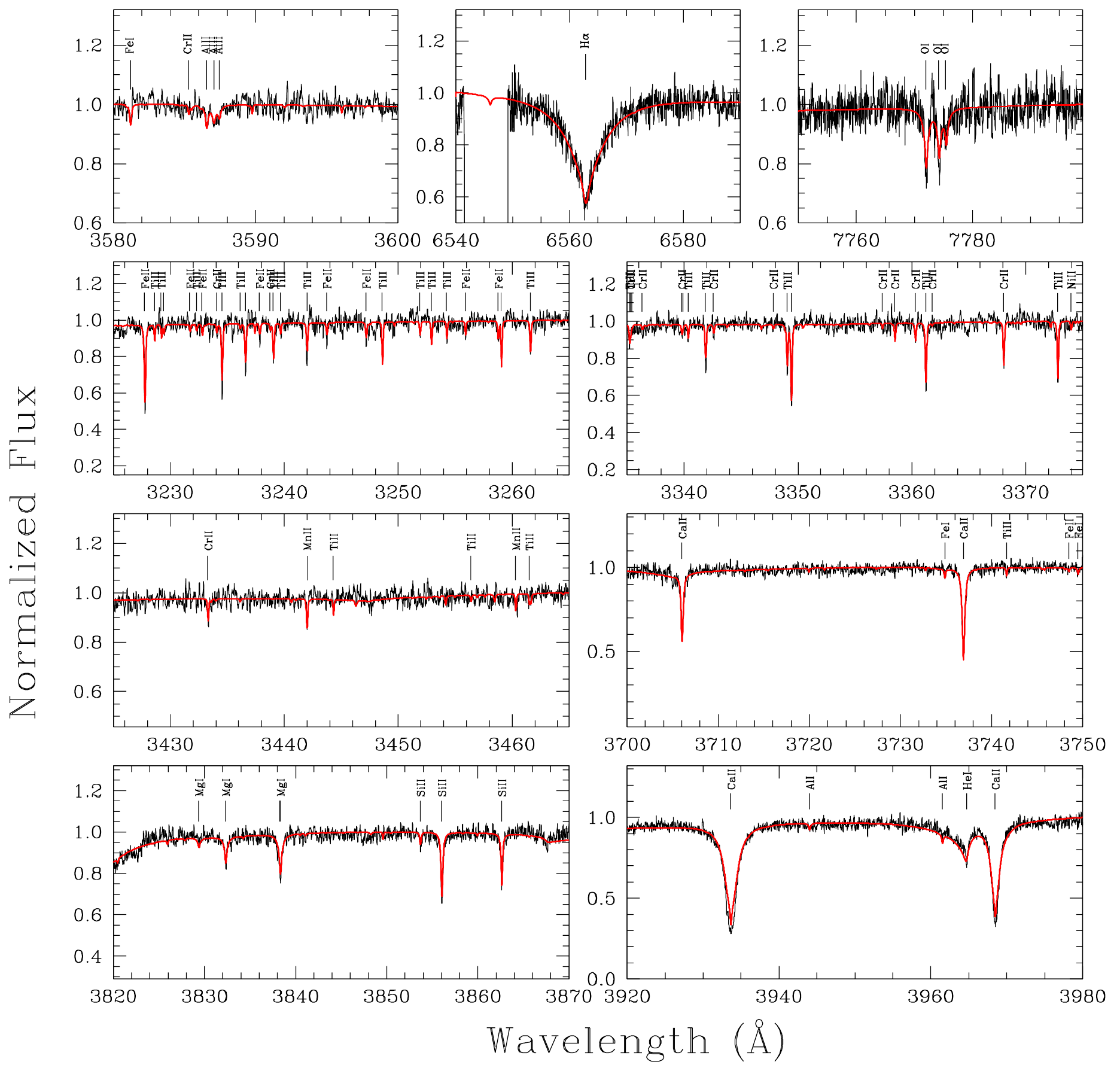}
\caption{Best fit model on the spectra of WD~1551+175.
\label{fig:Fit_WD1551+1753}}
\end{figure*}

\begin{figure*}
\includegraphics[width=\textwidth]{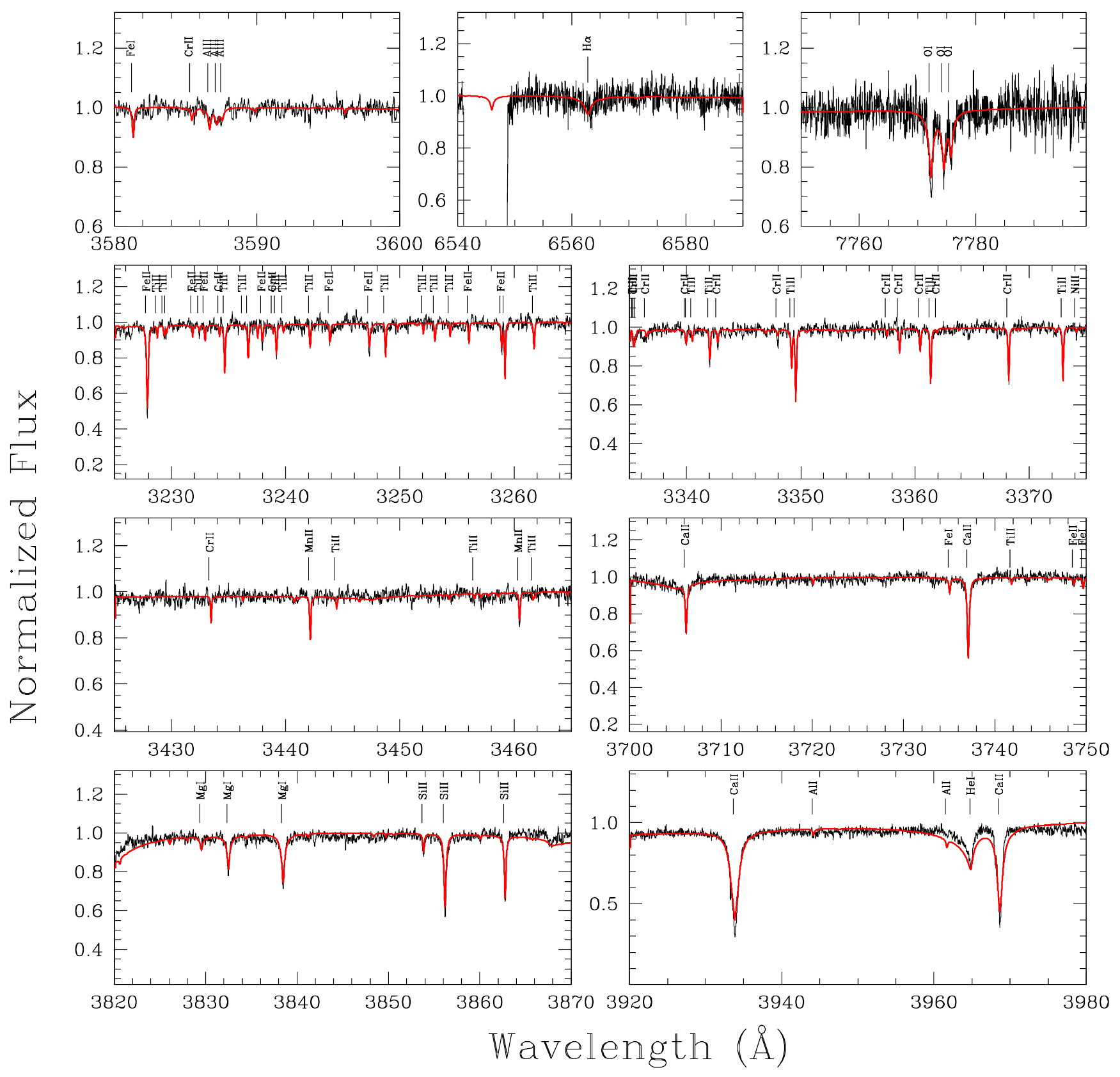}
\caption{Best fit model on the spectra of WD~2207+127.
\label{fig:Fit_WD2207+127}}
\end{figure*}

The observing log is listed in Table~\ref{tab:log} and some representative model fits to the data are shown in Figs.~\ref{fig:DAZCa} to \ref{fig:Fit_WD2207+127}.

\begin{deluxetable*}{llllllllll}
\tablecaption{Observing Logs}
\label{tab:log}
\tablewidth{0pt}
\tablehead{
Name & V	& Instrument & Resolution & Date (UT) &  Time & SNR\\
& (mag) && & & (sec)
}
\startdata
G 166-58	& 15.6 & HIRESb&	40,000& 2006 Jun 17	& 9600 & 80\\
&	& HIRESr	& 	40,000&2012 Dec 31	& 1800 & 30 \\
WD 2221-165	& 16.0 & HIRESb	&	40,000& 2012 Oct 29	& 2700 & 30\\
	& & HIRESr	&40,000 & 2012 Aug 02	& 3250 & 25\\
WD 0307+077 & 16.0 & HIRESr	&  40,000& 2012 Dec 31	& 3600 & 20\\
PG 1541+651	& 15.5 & HIRESb &40,000	& 2012 Apr 14 & 3000 & 25 \\
	& & HIRESr	& 40,000	& 2012 Aug 02	& 3600 & 40 \\
WD 1145+288	& 17.7 & HIRESb	& 40,000	& 2018 Jan 01	& 3000 & 10 \\
WD 1150-153	& 16.0 &HIRESb	&40,000 & 2008 Feb 13 & 3600 & 35\\
&	& HIRESr	& 40,000& 2012 Dec 31 & 3600 & 25 \\
GD 56	& 15.5 &HIRESb & 40,000	& 2007 Nov 20	& 2700& 30 \\
&		& HIRESr & 40,000		& 2008 Nov 14, 2019 Sept 7	& 3900 & 50 \\
WD 0107-192	& 16.2 &UVES	& 22,000	&  2015 Dec 26 & 3600 & 20\\
HE 0106-3253	& 15.4 & HIRESb	& 40,000& 2008 Aug 07	& 3000 & 50\\
&	& HIRESr	& 40,000& 2006 Sept 02	& 4800 & 60\\
PG 1015+161	&15.6 & HIRESb &	40,000& 2007 May 6, 2008 Feb 13 & 5400 & 65\\
	&& HIRESr	& 40,000& 2008 Feb 26 & 3300 & 50\\
PG 1457-086	& 15.8 &HIRESb	& 40,000& 2012 Apr 14 & 3600 & 35 \\
	& & HIRESr	& 40,000& 2012 Aug 02	& 3600 & 35\\
WD 1226+110	& 16.4 & HIRESb &40,000	& 2007 May 05 & 3000 & 35\\
	& & HIRESr	& 40,000& 2008 Nov 14,15,16	& 5300 & 10\\
PG 1018+411	& 16.4 & ESI & 14,000 	& 2017 Mar 07 & 3000 & 50\\
PG 0843+517	& 16.1 & HIRESb & 40,000& 2012 Jan 15  & 1800 & 20\\ 
&	& HIRESr	& 40,000& 2012 Dec 31	& 3600 & 25\\
WD 1341+036		& 17.0 & ESI &	14,000& 2017 Apr 17	& 2600 & 50\\
WD 0010+280 & 15.7 &COS & 18,000 & 2016 Jun 10 &	1761 & 15\\
\\
WD 1232+563	& 18.2 & HIRESb	&40,000	& 2015 Apr 11, 2016 Apr 01	& 10800& 20 \\
	& & ESI	& 14,000& 2015 Apr 25, 2017 Mar 06 & 8460 & 65\\
WD 1551+175	& 17.5 & HIRESb	& 40,000& 2013 May 08 & 4800 & 30\\
&	& HIRESr	& 40,000	&2015 Apr 09	& 6000 & 25 \\	
WD 2207+121	&17.3 & HIRESb	& 40,000& 2012 Oct 28,29	& 8600 & 45\\ 
	& & HIRESr	& 40,000& 2013 Sept 17	& 9600 & 25\\
\enddata
\tablecomments{SNR is measured around 3940~{\AA} for HIRESb and UVES, 6600~{\AA} for HIRESr and ESI, and 1350~{\AA} for COS.
}
\end{deluxetable*}

\end{CJK}

\clearpage
\bibliography{WD.bib}
\end{document}